\documentclass[preprintnumbers,article,amsmath,amssymb,floatfix,10pt,prd,twocolumn,
superscriptaddress,nofootinbib]{revtex4-2}
\usepackage{bm}
\usepackage{amsfonts}
\usepackage{latexsym}
\usepackage[latin1]{inputenc}
\usepackage{graphicx}
\usepackage{amsmath}
\usepackage{palatino}
\usepackage{mathpazo}
\usepackage{textcomp}
\usepackage{float}
\usepackage{booktabs}
\usepackage{dcolumn}
\usepackage{ragged2e}
\usepackage{hyperref}
\hypersetup{colorlinks,citecolor=blue}
\hypersetup{colorlinks=true,linkcolor=red,filecolor=magenta,    urlcolor=blue}
\usepackage{amsmath}
\usepackage{xcolor}
\usepackage{orcidlink}
\usepackage{epsfig}
\usepackage{caption}
\usepackage{subcaption}
\usepackage{commath}
\def\jnl@style{\it}
\def\aaref@jnl#1{{\jnl@style#1}}

\def\aaref@jnl#1{{\jnl@style#1}}

\def\aj{\aaref@jnl{AJ}}                   
\def\apj{\aaref@jnl{ApJ}}                 
\def\apjl{\aaref@jnl{ApJ}}                
\def\apjs{\aaref@jnl{ApJS}}               
\def\apss{\aaref@jnl{Ap\&SS}}             
\def\aap{\aaref@jnl{A\&A}}                
\def\aapr{\aaref@jnl{A\&A~Rev.}}          
\def\aaps{\aaref@jnl{A\&AS}}              
\def\mnras{\aaref@jnl{Mon.~Not.~Roy.~Astron.~Soc.}}             
\def\prd{\aaref@jnl{Phys.~Rev.~D}}        
\def\prc{\aaref@jnl{Phys.~Rev.~C}}  
\def\prl{\aaref@jnl{Phys.~Rev.~Lett.}}    
\def\qjras{\aaref@jnl{QJRAS}}             
\def\skytel{\aaref@jnl{S\&T}}             
\def\ssr{\aaref@jnl{Space~Sci.~Rev.}}     
\def\zap{\aaref@jnl{ZAp}}                 
\def\nat{\aaref@jnl{Nature}}              
\def\aplett{\aaref@jnl{Astrophys.~Lett.}} 
\def\apspr{\aaref@jnl{Astrophys.~Space~Phys.~Res.}} 
\def\physrep{\aaref@jnl{Phys.~Rep.}}      
\def\physscr{\aaref@jnl{Phys.~Scr}}       
\def\commat{\aaref@jnl{Comm.~Math.~Phys.}}              
\def\science{\aaref@jnl{Science}}               
\def\cqg{\aaref@jnl{Classical Quant.~Grav.}}            
\def\jpcs{\aaref@jnl{JPCS}}                                     
\def\ijmpd{\aaref@jnl{Int.~J.~Mod.~Phys.~D}}                    
\def\grg{\aaref@jnl{Gen.~Relat.~Gravit.}}               
\def\rpp{\aaref@jnl{Rep.~Prog.~Phys.}}          
\def\npa{\aaref@jnl{Nucl.~Phys.~A}}        
\def\lrr{\aaref@jnl{Living Rev.~Rel.}}                   
\def\jcap{\aaref@jnl{J.~Cosmology Astropart.~Phys.}}    
\def\rmp{\aaref@jnl{Rev.~Mod.~Phys.}}   
\def\epjc{\aaref@jnl{Eur.~Phys.~J.~C}} 
\def\plb{\aaref@jnl{~Phy.~Lett.~B}} 
\def\mpla{\aaref@jnl{Mod.~Phy.~Lett.~A}} 
\def\arxiv{\aaref@jnl{arxiv.org}}

\allowdisplaybreaks[1]

\begin{document}
\color{black}       
\title{Crossing phantom divide in $f(Q)$ gravity}

\author{Simran Arora\orcidlink{0000-0003-0326-8945}}
\email{dawrasimran27@gmail.com}
\affiliation{Department of Mathematics, Birla Institute of Technology and
Science-Pilani,\\ Hyderabad Campus, Hyderabad-500078, India.}

\author{P.K. Sahoo\orcidlink{0000-0003-2130-8832}}
\email{pksahoo@hyderabad.bits-pilani.ac.in}
\affiliation{Department of Mathematics, Birla Institute of Technology and
Science-Pilani,\\ Hyderabad Campus, Hyderabad-500078, India.}
%

\begin{abstract}
We investigate the possibility of crossing a phantom divide line in the extension of symmetric teleparallel gravity or the $f(Q)$ gravity, where $Q$ is the non-metricity. We study the cosmic evolution of the effective equation of state parameter for dark energy considering exponential, logarithmic, and combined $f(Q)$ theories. Moreover, the exponential model behaves like the $\Lambda$CDM at high redshifts before deviating to $\omega_{eff}<-1$ or $\omega_{eff}>-1$, respectively, depending on the value of model parameter. It also approaches a de-sitter phase asymptotically. However, the crossing of the phantom divide line, i.e., $\omega= -1$, is realized in the combined $f(Q)$ theory. Furthermore, statefinder diagnostics are studied in order to differentiate between several dark energy models. To ensure the three model's stability, we employ the stability analysis  using linear perturbations. We demonstrate how to reassemble $f(Q)$ via a numerical inversion approach based on existing observational constraints on cosmographic parameters and the potential of bridging the phantom divide in the resulting model. It explicitly demonstrates that future crossings of the phantom dividing line are a generic feature of feasible $f(Q)$ gravity models.

\end{abstract}


\maketitle

\date{\today}

\section{Introduction}

The primary purpose of current and future cosmological investigations is to learn more about the true nature of cosmic acceleration by subjecting the conventional cosmological model $\Lambda$CDM ($\Lambda$ cold dark matter) to the test as well as variations from it \cite{Riess/1998,Perlmutter/1999,Ade/2016,Aghanim/2020}. Using the cosmological constant as the fundamental source of the accelerated behavior, one can develop gravity theories that are equivalent to $\Lambda$CDM at the background level yet exhibit distinct and fascinating signatures on perturbation dynamics. Following this notion, we will see if there is a gravity theory with these characteristics that can confront the $\Lambda$CDM paradigm. \\
One of the simplest possibility is introducing an arbitrary function of the Ricci scalar $R$ in the action, which gives the $f(R)$ theory \cite{Buchdahl/1970,Capo/2008,Cruz/2006}.  Another two approaches after curvature representations are: the teleparallel gravity in which gravitational force is governed by the torsion $T$ \cite{Capo/2011,Liu/2012,Iorio/2012,Wang/2020,Nunes/2016}. Another possibility employed by Einstein is symmetric teleparallel gravity which attempts to unify field theories. It accounts vanishing curvature and torsion with non-vanishing non-metricity. We will focus at modified gravity based on non-metricity which is a quantity that analyzes how the length of a vector changes when transported. \\
The symmetric teleparallel gravity or the $f(Q)$ gravity \cite{Nester/1999,Lazkoz/2019,Jimenez/2018,Jimenez/2020} is a modified gravity that has recently sparked a lot of interest. When pertubations around the FLRW backgorund are addressed, the theory shows no strong coupling issues. The $f(Q)$ theory has been studied in a variety of ways, including observational data constraints for various parametrizations of $f(Q)$ \cite{Lazkoz/2019,Ayuso/2021}, cosmography \cite{Mandal/2020}, energy conditions \cite{Mandal/2020a}, black holes \cite{Ambrosio/2022}, bouncing \cite{Bajardi/2020, Mandal/2021},  evolution of the growth index in matter perturbations and more \cite{Khyllep/2021}. For more works in $f(Q)$ theory, one can check \cite{Albuquerque/2022,Esposito/2022,Atayde/2021,Zhao/2022,Frusciante/2021,Dimakis/2021,Barros/2020,Harko/2018,Bajardi/2020}. \\
The crossing of the phantom divide line by the dark energy equation of state has two possible cosmic indications \cite{Apos/2006,Wu/2011,Nozari/2008,Bamba/2010,Zhao/2019,Karimzadeh/2019}. One is either the dark energy consists of numerous components, at least one of which is a non-canonical phantom component, or general relativity must be extended to a cosmological scale theory. The current study looks into the cosmic evolution in the exponential \cite{Yang/2010,Bamba/2011,Linder/2010} and logarithmic \cite{Bamba/2011} $f(Q)$ theories. We investigate the energy density and equation of state parameter in the context of dark energy. The universe  with exponential gravity in $f(Q)$ stays in either a non-phantom or a phantom phase without crossing the phantom division line as shown in further study. Moreover, the model behaves like the $\Lambda$CDM at high redshifts before deviating to $\omega_{eff}<-1$ or $\omega_{eff}>-1$, respectively depending on the value of a parameter in a model. It also approaches a de-sitter phase asymptotically. Also, the logarithmic $f(Q)$ model behaves in a similar fashion. To realise the crossing of phantom divide line, we use a composite $f(Q)$ model with exponential and logarithmic factor. In this case, $\omega_{eff}$ can be seen crossing the phantom division line.\\

With a negative equation of state, one may deduce the existence of dark energy, but inferring the information about the dynamical feature of $\omega$ is difficult. It appears that a higher time derivative of the scale factor is required to obtain the information on the dynamical evolution of $\omega$. The virtue of the statefinder parameters features all kinds of dark energy models. In fact, the statefinder diagnostic has been widely used in a variety of models, including $\Lambda$CDM, Chaplygin gas \cite{Alam/2003,Gorini/2003}, quintessence \cite{Sahni/2003,Zhang/2005}, holographic dark energy models \cite{Zhang/2008,Setare/2007}, phantom models \cite{Chang/2007}, etc.
The spatially flat $\Lambda$CDM model corresponds to a fixed point $(r,s)=(1,0)$. The deviation of a specific dark energy model from this fixed point is a suitable approach to determine the model's departure from the flat $\Lambda$CDM.\\
In addition, we are interested in investigating the stability of cosmological solutions in non-metric models when perturbed by homogeneous perturbations. Typically, homogeneous and isotropic perturbations have been used to determine the stability of several modified gravity theories \cite{Farrugia/2016,Dombriz/2012}.\\
The next section introduces $f(Q)$ gravity and derives the field equations. The cosmological formulation of the $f(Q)$ theory is discussed in section \ref{section 3}. The cosmic evolution of exponential, logarithmic, and combined $f(Q)$ theories is investigated in section \ref{section 4}. In section \ref{section 5}, we discussed the statefinder analysis followed by the linear perturbation analysis in section \ref{section 6}. We also reconstructed the $F(Q)$ gravity using cosmography in section \ref{section 7}. Our conclusions are discussed in section \ref{section 8}.

\section{Overview of $f(Q)$ gravity}
\label{section 2}

The action for the non-metricity based $f(Q)$ gravity \cite{Jimenez/2018} is given as follows:

\begin{equation} \label{1}
S= \int \left[ \frac{1}{2} f(Q)+L_{m} \right] \sqrt{-g} d^{4}x,
\end{equation}
where $L_{m}$ represents the matter Lagrangian, $g$ is the determinant of the metric $g_{\mu \nu}$, and $f(Q)$ is an arbitrary function of $Q$. The basic object in this class of theories is the non-metricity tensor defined as $Q_{\alpha\mu\nu}= \nabla_{\alpha}g_{\mu \nu}$. 
Further, the two independent traces and the non-metricity scalar are defined as $Q_{\alpha} = g^{\mu \nu} Q_{\alpha \mu \nu}$,  $\overline{Q}_{\alpha} =g^{\mu \nu} Q_{ \mu \alpha \nu}$ and $Q = -Q_{\alpha \mu \nu} P^{\alpha \mu \nu}$. \\
The latter expression includes the non-metricity conjugate reads 
\begin{equation}\label{2}
P^{\alpha}_{\mu \nu}= -\frac{1}{2} L^{\alpha}_{\mu \nu} + \frac{1}{4} (Q^{\alpha}-\overline{Q}^{\alpha})g_{\mu \nu}-\frac{1}{4} \delta^{\alpha}_{(\mu}Q_{\nu)}, 
\end{equation}
 and the disformation tensor
\begin{equation} \label{3}
L^{\alpha}_{\mu \nu}= \frac{1}{2} Q^{\alpha}_{\mu \nu}-Q_{(\mu\nu)}^{\alpha}.
\end{equation} 

Let us recall that for $f(Q)=-Q$ \cite{Jimenez/2018}, the action \eqref{1} has been shown to be identical to general relativity (GR) in flat space. As a result, any alteration from GR can be turned into $f(Q)$.\\
The field equations describing the gravitational interaction in $f(Q)$ gravity is obtained by varying the action with respect to metric tensor 
\begin{multline}  \label{4}
\frac{2}{\sqrt{-g}} \nabla_{\lambda}(\sqrt{-g} f_{Q} P^{\lambda}_{\mu \nu} )+\frac{1}{2} g_{\mu \nu} f + f_{Q} (P_{\mu \lambda \beta} Q_{\nu}^{\lambda \beta}\\
 - 2Q_{\lambda \beta \mu} P^{\lambda \beta}_{\nu})= -T_{\mu \nu},
\end{multline}
where $f_{Q}=\frac{df}{dQ}$. Here, $T_{\mu \nu}$ is the stress-energy momentum tensor  given by 
\begin{equation} \label{5}
T_{\mu \nu}= \frac{-2}{\sqrt{-g}} \frac{\delta(\sqrt{-g} L_{m})}{\delta g^{\mu \nu}}.
\end{equation}

\section{Cosmological formulation}
\label{section 3}
We start with the Friedmann-Lema\^{i}tre-Robertson-Walker (FLRW) metric in a homogeneous and isotropic flat spacetime, 
\begin{equation} \label{6}
ds^{2}= -dt^{2}+a^{2}(t)[dx^{2}+dy^{2}+dz^{2}],
\end{equation}
where $a(t)$ is the scale factor of the universe.
One can obtain the non-metricity scalar as $Q=6 H^{2}$, where $H=\frac{\dot{a}}{a}$ is the Hubble parameter. The energy-momentum tensor $T_{\mu \nu}$ for the perfect fluid is given by 
\begin{equation}\label{7}
T_{\mu \nu}= (\rho+p) u_{\mu}u_{\nu}+pg_{\mu \nu},
\end{equation}
where $u_{\mu}$ is the four-velocity satisfying the condition $u_{\mu} u^{\mu}=-1$, $\rho$ and $p$ are the energy density and pressure of a perfect fluid respectively.

On taking $f(Q)=-Q+F(Q)$, the corresponding field equations are 
\begin{equation} \label{8}
H^{2}= \frac{1}{3} \rho_{m} -\frac{F}{6}+\frac{1}{3} Q F_{Q},
\end{equation}
\begin{equation} \label{9}
(H^{2})'= \frac{2p_{m}+F+Q-2QF_{Q}}{2(2QF_{QQ}+F_{Q}-1)},
\end{equation}
where prime denotes a derivative with respect to $ln~a$, $F_{Q}=\frac{dF(Q)}{dQ}$, $F_{QQ}=\frac{d^{2}F(Q)}{dQ^{2}}$. Here, $\rho_{m}$, $p_{m}$ are the energy density and pressure of generic matter, respectively.

By comparing the above modified Friedmann equations with the ordinary ones in general relativity
\begin{equation} \label{10}
H^{2}= \frac{1}{3}(\rho_{m}+\rho_{eff}), 
\end{equation}
\begin{equation} \label{11}
(H^{2})'= - (\rho_{m}+ p_{m}+\rho_{eff}+p_{eff}),
\end{equation}
we get, 
\begin{equation} \label{12}
\rho_{eff}= \frac{1}{2}(-F+ 2 Q F_{Q}),
\end{equation}
\begin{equation} \label{13}
p_{eff}= \frac{F/Q - F_{Q}+QF_{QQ}}{(2QF_{QQ}+F_{Q}-1)(F/Q - 2F_{Q})}.
\end{equation}
Further, using eqns. \eqref{12} and \eqref{13}, the equation of state parameter is obtained as
\begin{align} \label{14}
\omega_{eff} =& -1 +4\dot{H} \left( \frac{F_{Q}+2QF_{QQ}}{F-2QF_{Q}}\right) \\
=& \frac{F/Q -F_{Q}+2QF_{QQ}}{(F/Q -2F_{Q})(2QF_{QQ}+F_{Q}-1)}.
\end{align}

Considering the non-relativistic matter (i.e., $\rho_{m}$ and $p_{m}=0$), the effective dark energy can be seen satisfying the continuity equation:
\begin{equation} \label{16}
\rho_{eff}'= \frac{d\rho_{eff}}{d ln~a}= -3 (1+\omega_{eff})\rho_{eff}. 
\end{equation}

\section{cosmological evolution}
\label{section 4}

We define the dimensionless parameter to study the cosmic evolution of the effective equation of state in $f(Q)$ theory \cite{Hu/2007,Bamba/2011}
\begin{equation} \label{17}
x_{H}= \frac{H^{2}}{\overline{m}^{2}}-a^{-3}= \frac{\rho_{eff}}{\rho_{m}^{0}},
\end{equation}
where $\overline{m}^{2}= \frac{8 \pi G \rho_{m}^{0}}{3}$. Here $\rho_{m}^{0}= \rho_{m} (z=0)$ is the present density parameter of non-relativistic matter. From \eqref{16}, we have the following:
\begin{equation} \label{18}
x_{H}'= -3(1+\omega_{eff})x_{H}. 
\end{equation}
As a result of \eqref{17}, we obtained $H^{2}= \overline{m}^{2}(x_{H}+a^{-3})$. It is important to notice that $\omega_{eff}$ is a function of $Q$, whereas $Q$ is a function of $H^{2}$. The relation $a= \frac{1}{1+z}$ is used in subsequent calculations.\\

\subsection{Model I}

As a first, we consider the exponential $F(Q)$ theory given by \cite{Linder/2010,Bamba/2011}
\begin{equation}
\label{19}
F(Q)= \alpha Q (1-e^{s Q_{0}/Q}),
\end{equation}
with $\alpha= \frac{1-\Omega_{m}^{0}}{1-(1-2s)e^{s}}$,  and a constant  $s$. In particular, $s=0$ corresponds to the usual Friedmann equations as expected. This particular functional form corresponds to the symmetric teleparallel equivalent of general relativity (STEGR).
Here, $\Omega_{m}^{0}= \frac{\rho_{m}^{0}}{\rho_{c}^{0}}$, where $\rho_{m}^{0}$ is the energy density of non-relativistic matter at the present time and $\rho_{c}^{0}= \frac{3H_{0}^{2}}{8 \pi G}$ is the critical density. Note that the parameters with superscript ``0" represents the value at $z=0$.\\
From the above relations, the dimensionless quantities at $z=0$ are calculated as:
$\frac{\overline{m}^{2}}{Q_{0}}= \frac{\Omega_{m}^{0}}{6}$ and $x_{H}(z=0)=\frac{1}{\Omega_{m}^{0}}-1=2.17$ for $\Omega_{m}^{0}=0.315$ \cite{Aghanim/2020}.

\begin{widetext}

\begin{figure}[H]
\centering
\begin{subfigure}{0.45\textwidth}
    \includegraphics[width=\textwidth]{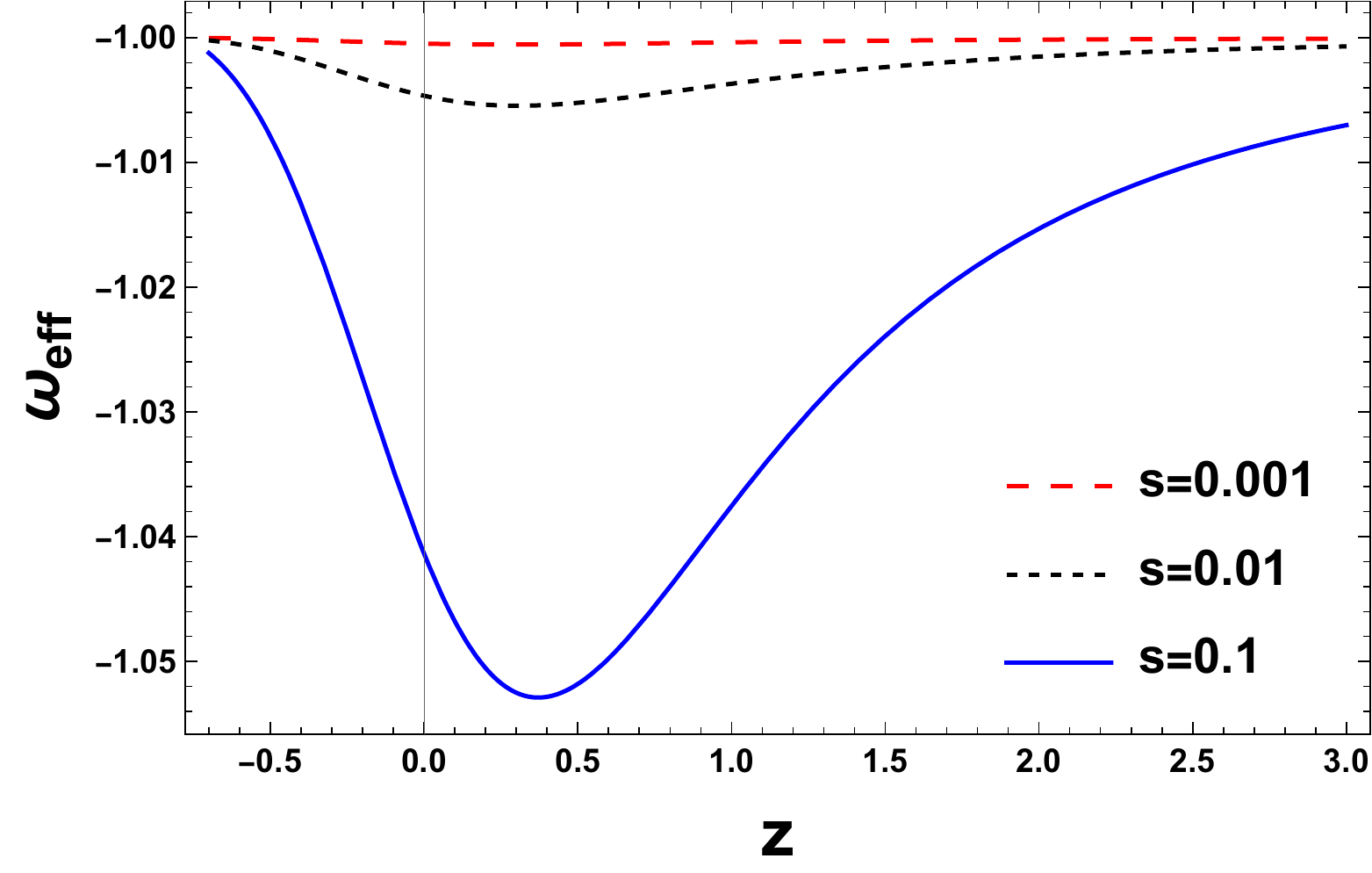}
    \caption{}
    \label{W+}
\end{subfigure}
\hfill
\begin{subfigure}{0.45\textwidth}
    \includegraphics[width=\textwidth]{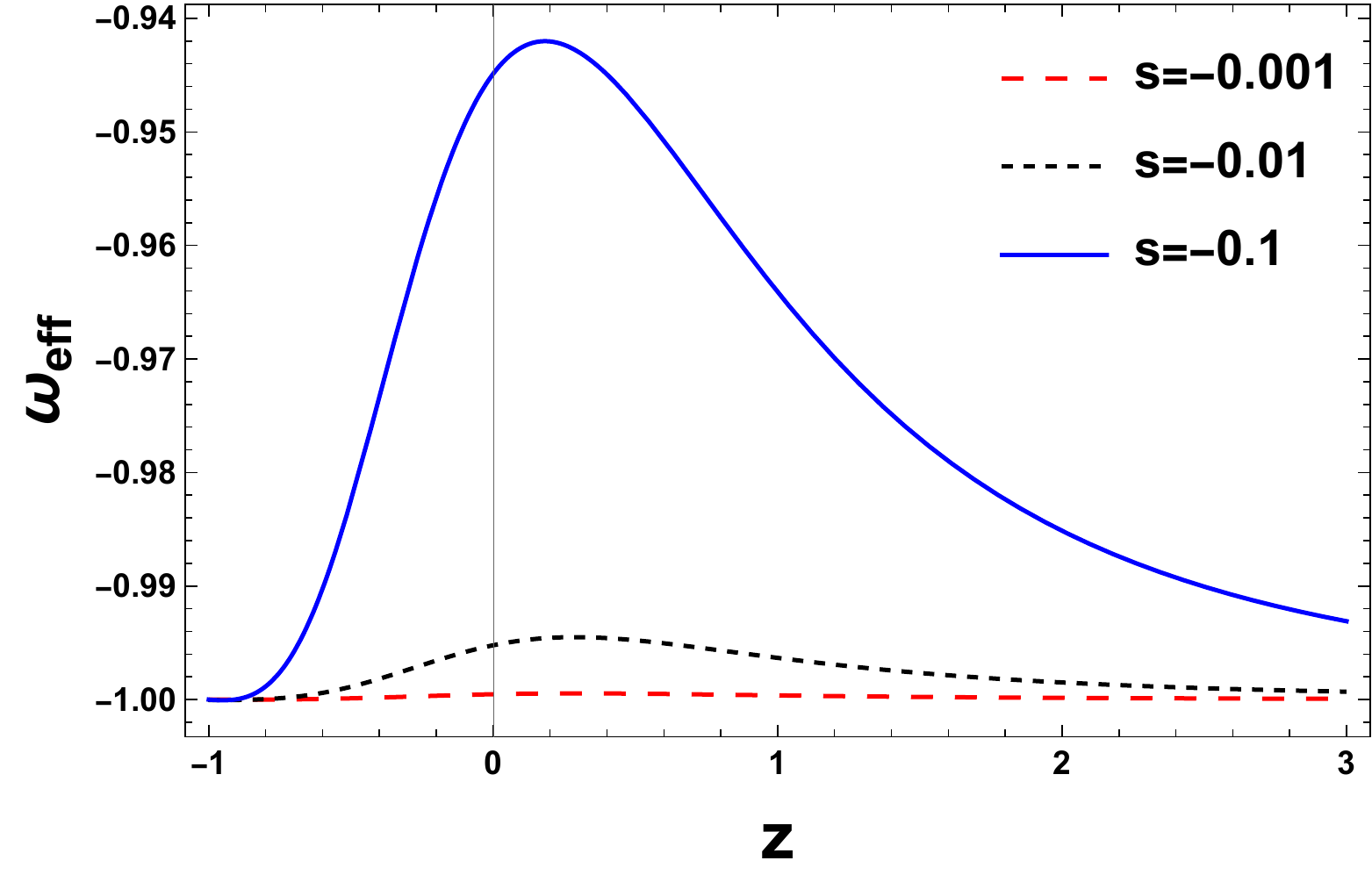}
    \caption{}
    \label{W-}
\end{subfigure}
\caption{Evolution of $\omega_{eff}$ as a function of redshift $z$ for $|s|=0.001$, $0.01$, and $0.1$.}
\label{fig:1}
\end{figure}

\end{widetext}

\begin{widetext}

\begin{figure}[H]
\centering
\begin{subfigure}{0.45\textwidth}
    \includegraphics[width=\textwidth]{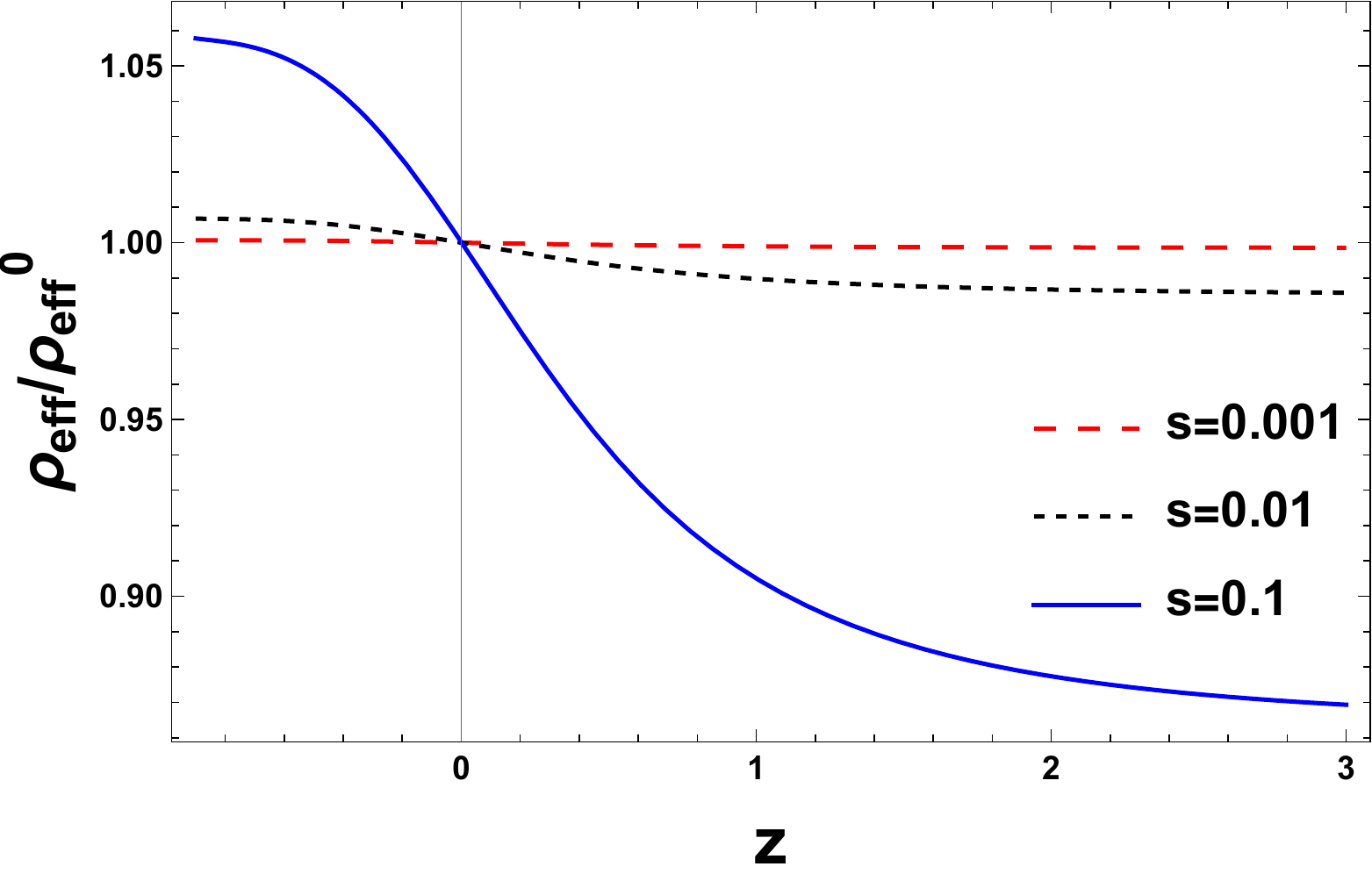}
    \caption{}
    \label{rho+}
\end{subfigure}
\hfill
\begin{subfigure}{0.45\textwidth}
    \includegraphics[width=\textwidth]{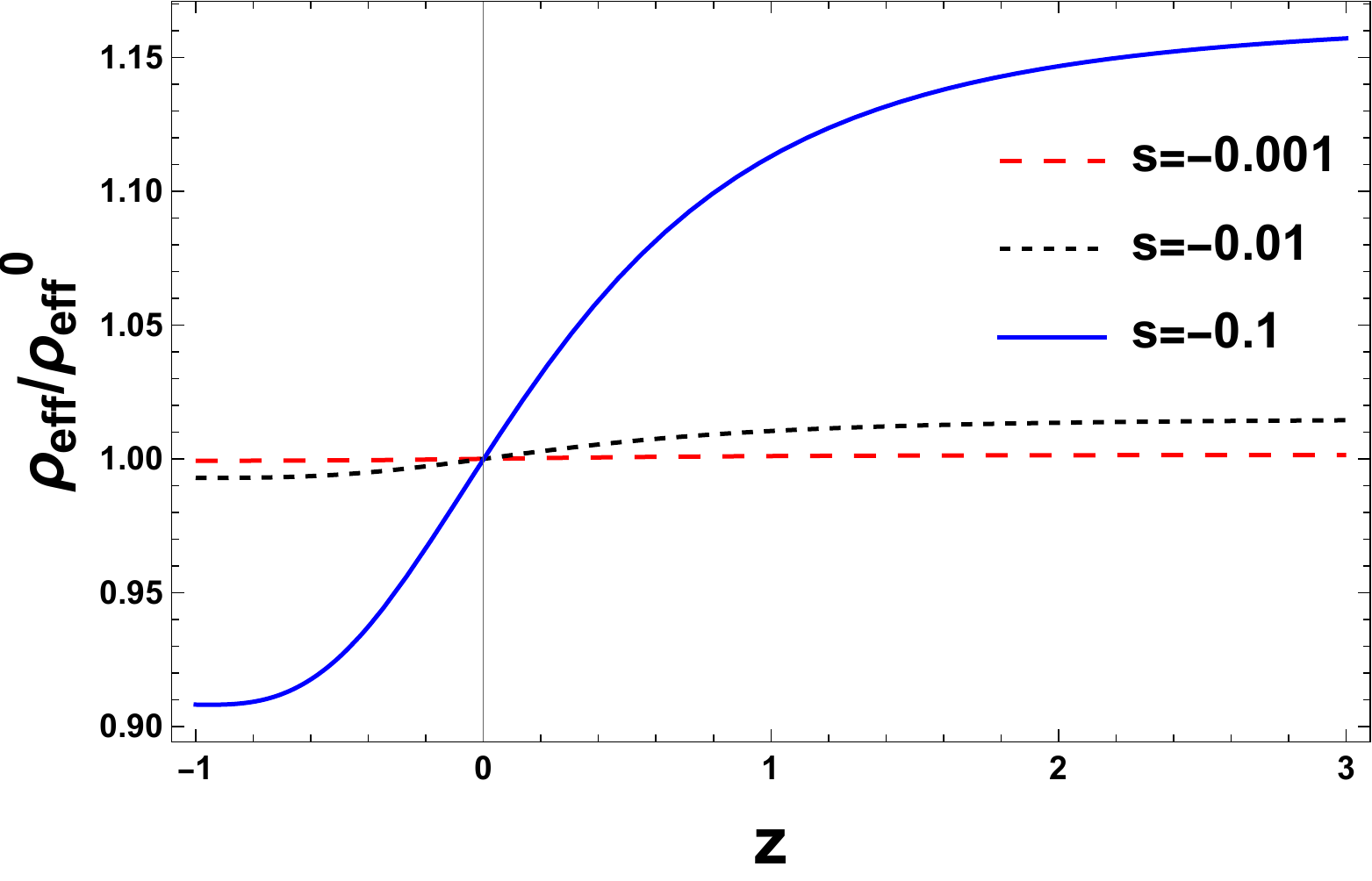}
    \caption{}
    \label{rho-}
\end{subfigure}
        
\caption{Evolution of $\rho_{eff}/\rho_{eff}^{0}$ as a function of redshift $z$ for $|s|=0.001$, $0.01$, and $0.1$.}

\end{figure}

\end{widetext}

The evolution of the equation of state parameter as functions of redshift for negative and positive values of parameter $s$ is presented  numerically in figs. \ref{W+} and \ref{W-}. It is observed that $\omega_{eff}$ does not cross the phantom divide line $\omega_{eff}=-1$ \cite{Linder/2010,Bamba/2011}. That is, it lies in  phantom phase for $s>0$ and lies in quintessence phase for $s<0$. In this work, we choose $|s|=0.001, 0.01, 0.1$ to see the behavior of exponential $f(Q)$ gravity. It is clear that deviation from $\Lambda$CDM model for exponential $f(Q)$ gravity is small for smaller $s$ and and large for  larger $s$. The present values of $\omega_{eff}^{0}$ are $-1.0004$, $-1.0046$, $-1.0413$, $-0.999$, $-0.995$, and $-0.944$ for $s=0.001$, $0.01$, $0.1$, $-0.001$, $-0.01$, and, $-0.1$, respectively \cite{Aghanim/2020,Bamba/2011,Linder/2010}.\\
Further figs. \ref{rho+} and \ref{rho-} depicts the evolution of $\rho_{eff}/\rho_{eff}^{0}$ versus redshift $z$. It is seen that $\rho_{eff}/\rho_{eff}^{0}$ becomes constant in the near future i.e. $z<0$.

\begin{widetext}

\begin{figure}[H]
\centering
\begin{subfigure}{0.45\textwidth}
    \includegraphics[width=\textwidth]{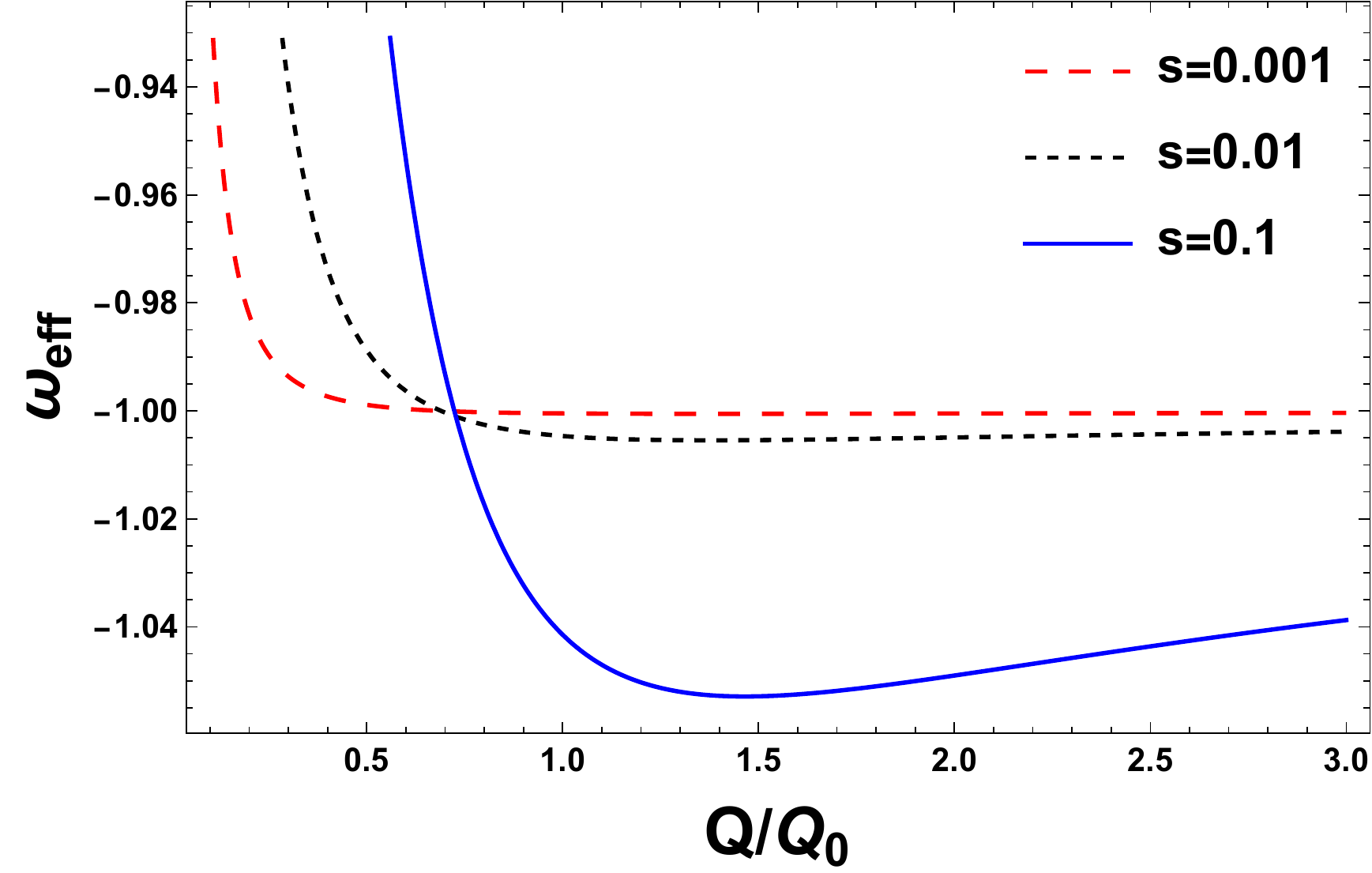}
    \caption{}
    \label{wQ+}
\end{subfigure}
\hfill
\begin{subfigure}{0.45\textwidth}
    \includegraphics[width=\textwidth]{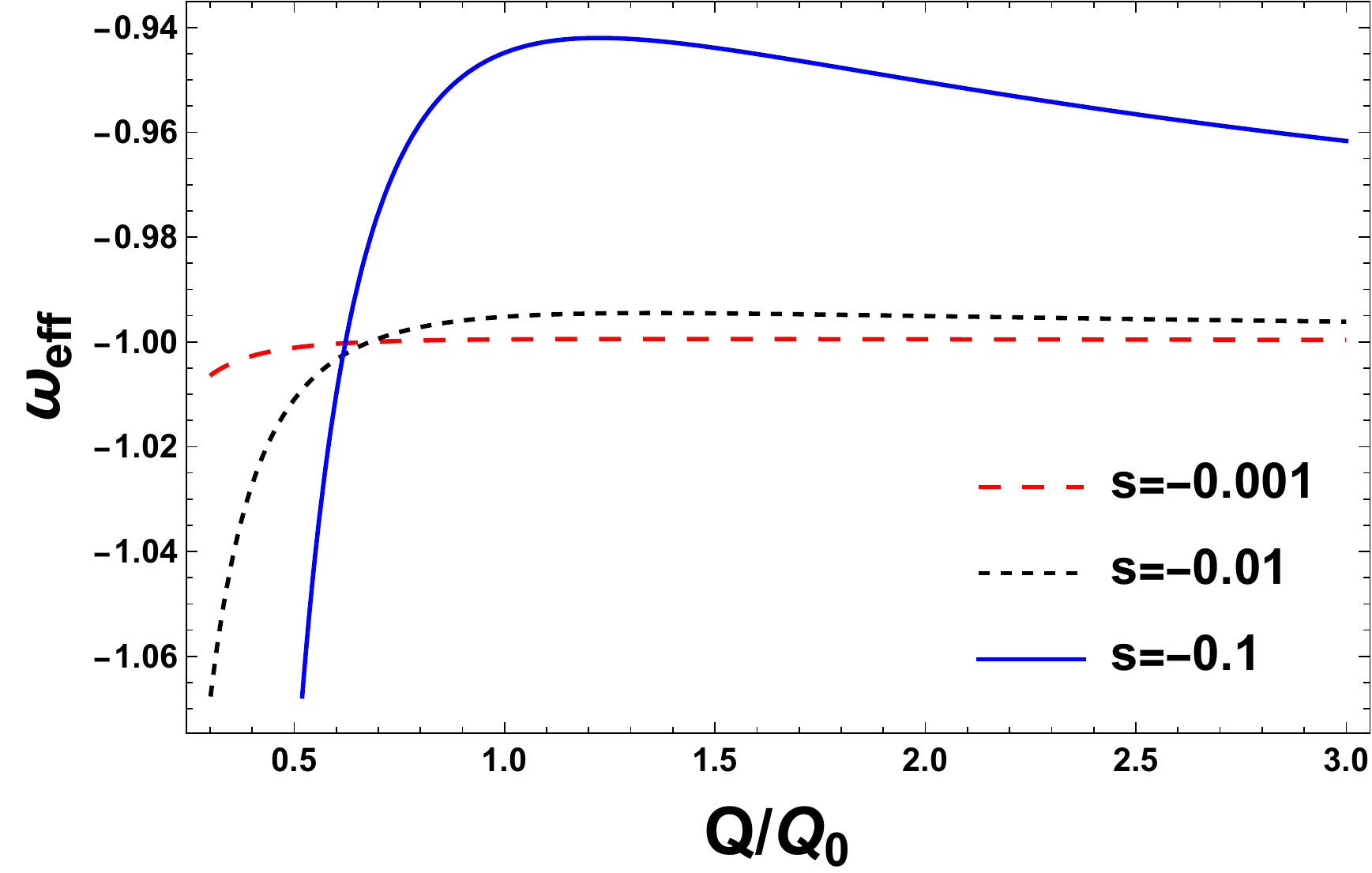}
    \caption{}
    \label{wQ-}
\end{subfigure}
\hfill

\caption{Evolution of $\omega_{eff}$ as a function of $Q/Q_{0}$ for $|s|=0.001$, $0.01$, and $0.1$.}
\label{fig:2}
\end{figure}

\end{widetext}

Figs. \ref{wQ+} and \ref{wQ-} illustrates the equation of state $\omega_{eff}$ as functions of $Q/Q_{0}$ for $s>0$ and $s<0$ respectively. It is clear that $\omega_{eff}$ versus $Q/Q_{0}$ crosses the phantom divide around $Q/Q_{0}$ with the large deviation from $\Lambda$CDM for $|s|=0.1$.  Furthermore, in cosmological context, the universe reaches the $\omega_{eff}=-1$ with the crossing point as $Q/Q_{0} \simeq 0.686$, $0.689$, $0.723$, $0.684$, $0.680$, $0.622$ for $s=0.001$, $0.01$, $0.1$, $-0.001$, $-0.01$, and, $-0.1$, respectively .\\

\textbf{Crossing of the phantom divide line:}\\

Note that from eq. \eqref{14}, $\omega_{eff}$ crosses $-1$ if $4\dot{H} \left( \frac{F_{Q}+2QF_{QQ}}{F-2QF_{Q}} \right)$ changes its sign.
According to eq. \eqref{19}, we have 
\begin{equation}
F_{Q}= \alpha \left( 1- e^{sQ_{0}/Q} + \frac{sQ_{0}}{Q} e^{sQ_{0}/Q} \right), 
\label{20}
\end{equation}
and 
\begin{equation}
F_{QQ}= -\alpha \left( \frac{sQ_{0}}{Q}\right) ^{2} \frac{1}{Q} e^{sQ_{0}/Q}. \label{21}
\end{equation}

\begin{widetext}

\begin{figure}[H]
\centering
\begin{subfigure}{0.45\textwidth}
    \includegraphics[width=\textwidth]{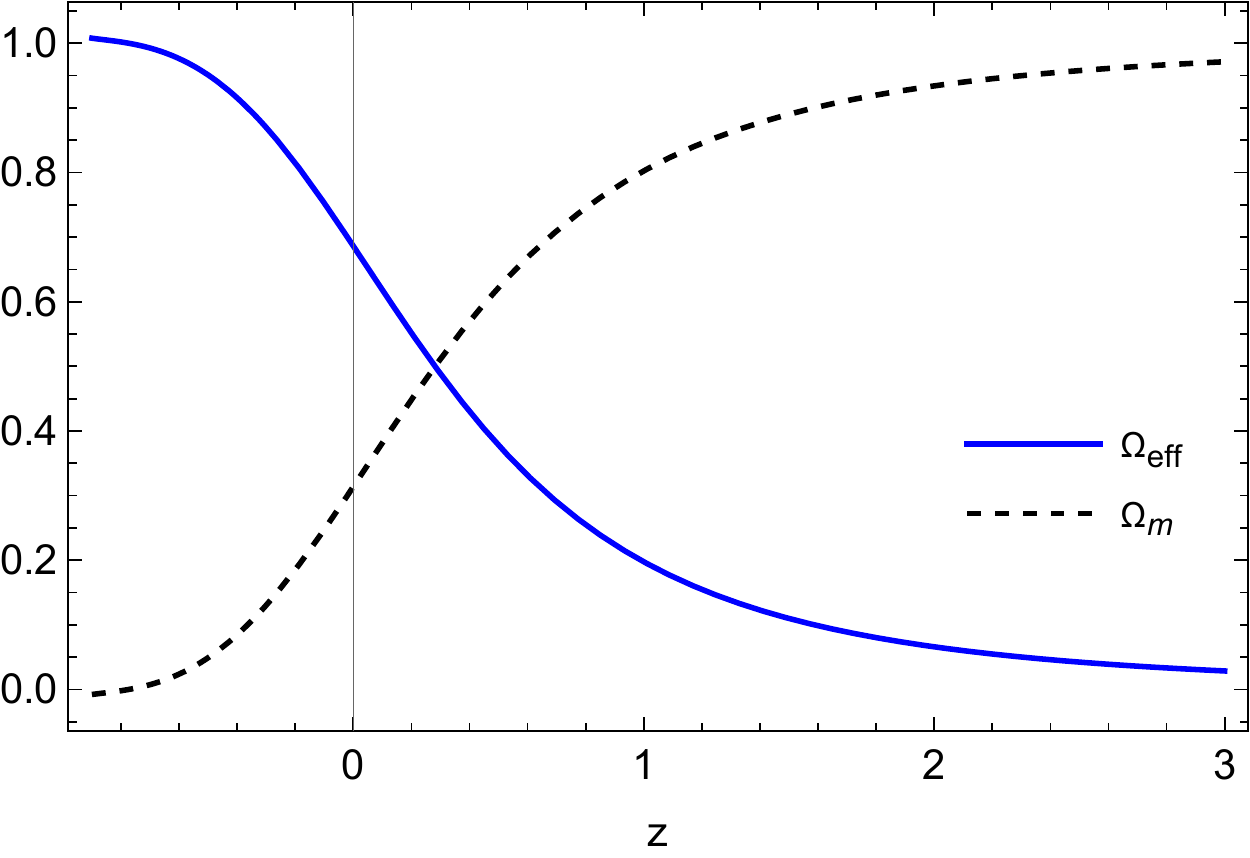}
    \caption{}
    \label{om+ EXP}
\end{subfigure}
\hfill
\begin{subfigure}{0.45\textwidth}
    \includegraphics[width=\textwidth]{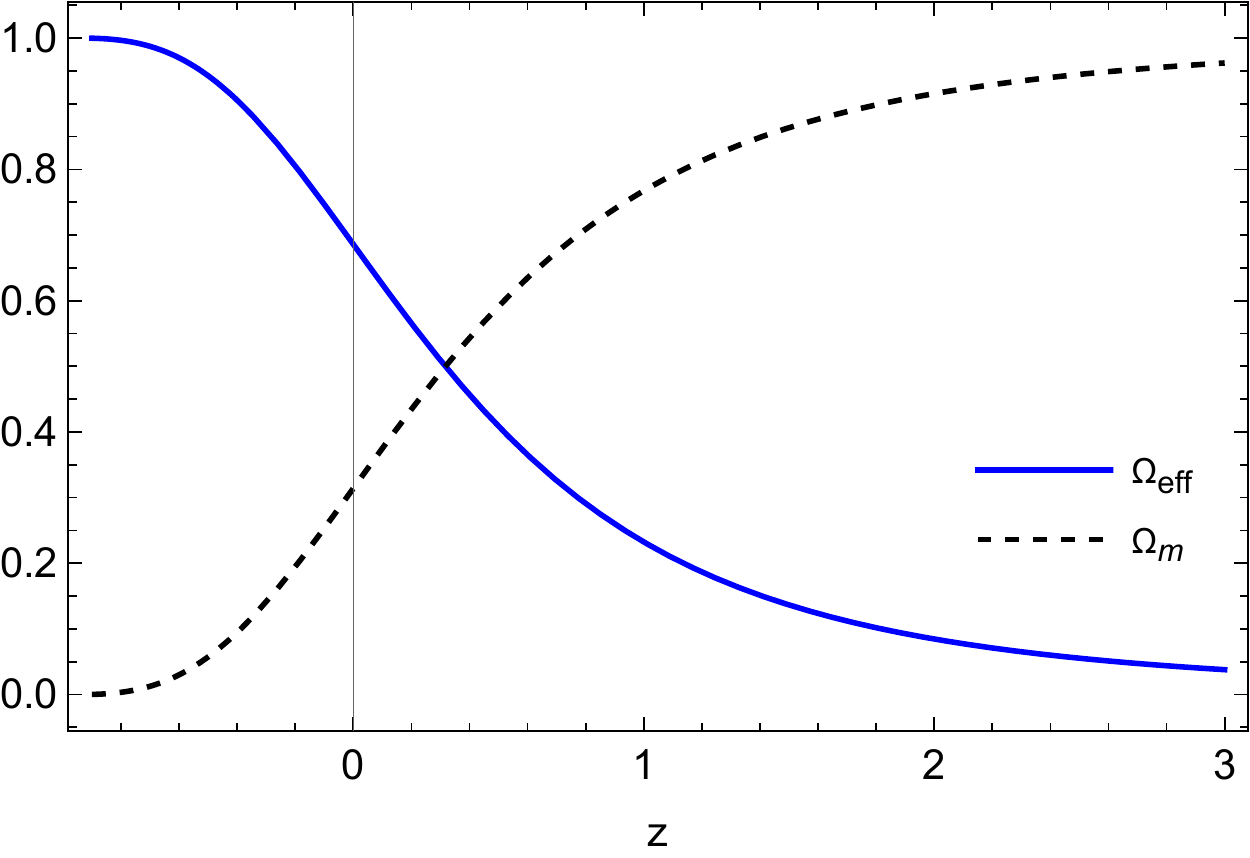}
    \caption{}
    \label{om- exp}
\end{subfigure}
\hfill

\caption{Evolution of $\Omega_{eff}$ and $\Omega_{m}$ as a function of $z$ for $|s|=0.1$ respectively.}
\label{fig:5}
\end{figure}

\end{widetext}

For the investigation, if we concentrate on $0 < s \ll1 $ and $A= \frac{sQ_{0}}{Q} \ll 1$. So, we obtained the following using approximation to second order:
\begin{equation}
\frac{F}{Q} -2 F_{Q} \approx -\alpha A \left( 1+ \frac{3A}{2} \right),
\end{equation}
and 
\begin{equation}
F_{Q}+2QF_{QQ} \approx -\frac{3 \alpha A^{2}}{2}.
\end{equation}
Here, both the equations  have same sign implies that $\frac{F_{Q}+2QF_{QQ}}{F/Q-2F_{Q}}$ does not change its sign. Henceforth, $\omega_{eff}$ do not cross the phantom divide line. \\
In figs. \ref{om+ EXP}, \ref{om- exp} we show the fractional densities of effective dark energy and the non-relativistic matter as functions of redshift $z$, i.e. $\left( \Omega_{eff}= \frac{\rho_{eff}}{\rho_{c}}, \Omega_{m}= \frac{\rho_{m}}{\rho_{c}}\right)$. Furthermore, the cosmological evolution of $\Omega_{eff}$ and $\Omega_{m}$ for other values of model parameter $|s|=0.01$ and $0.001$ is similar to $|s|=0.1$. The intersection of $\Omega_{eff}$ and $\Omega_{m}$ is the crossover point $z_{eff}$ at which $\Omega_{eff}=\Omega_{m}$. The universe reaches the matter-dominated stage as $z$ decreases. When $z<z_{eff}$, dark energy eventually overcomes matter. The present values for the parameters are obtained as $\Omega_{eff}^{(0)}=0.68$, $\Omega_{m}^{(0)}=0.314$, and $z_{eff}=0.28$ and $0.31$, respectively.

\subsection{Model II}
Here, we consider another motivated logarithmic $F(Q)$ model given by
\begin{equation}
F(Q)= \gamma Q_{0} \left( \frac{k Q_{0}}{Q}\right)^{-\frac{1}{2}} ln\left( \frac{k Q_{0}}{Q} \right),  
\end{equation} 
 
where $\gamma= \frac{\Omega_{m}^{0} -1}{2 k^{-1/2}}$ and a constant $k$. It is seen that the model contains only one parameter $k$ for the given $\Omega_{m}^{0}$.

\begin{figure}[]
\centering
\begin{subfigure}{0.45\textwidth}
    \includegraphics[width=\textwidth]{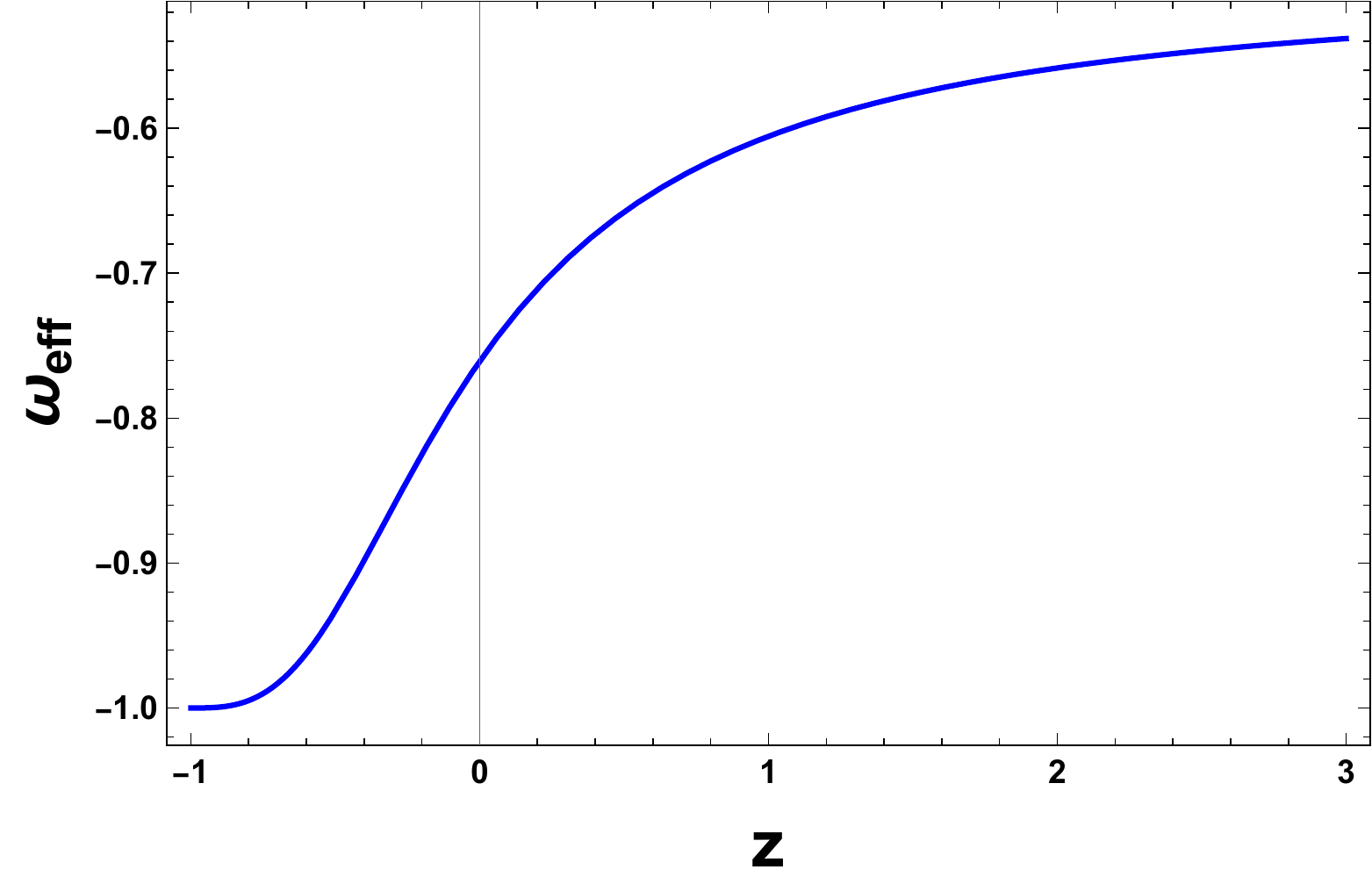}
    \caption{}
    \label{W+log}
\end{subfigure}
\hfill
\begin{subfigure}{0.45\textwidth}
    \includegraphics[width=\textwidth]{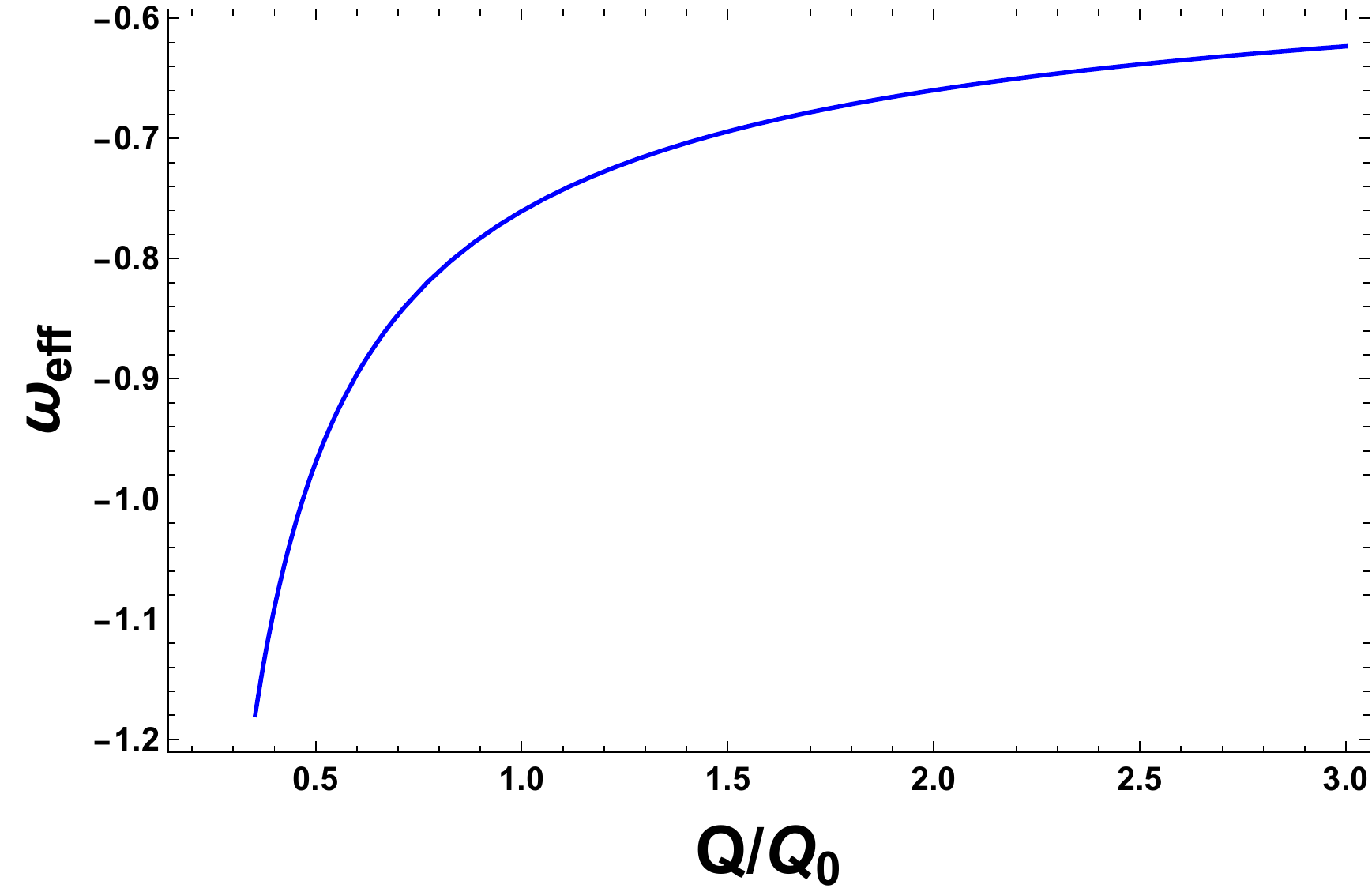}
    \caption{}
    \label{WQ+log}
\end{subfigure}
\hfill

\caption{Evolution of $\omega_{eff}$ as a function of redshift $z$ and $Q/Q_{0}$, respectively for $k=1$.}
\label{fig:3}
\end{figure}

Fig. \ref{W+log} and \ref{WQ+log} describes the behavior of $\omega_{eff}$ as functions of redshift $z$ and $Q/Q_{0}$ repectively for the parameter $k=1$. It is observed that $\omega_{eff}$ lies in a quintessence region and do not cross the phantom divide line. The present value of $\omega_{eff}^{0}$ is $-0.760$ \cite{Sanjay/2021}. After simplifying \eqref{14} for this model, we get
\begin{equation}
\omega_{eff}= - \frac{1}{2-(1-\Omega_{m}^{0}) \left( \frac{Q_{0}}{Q}\right)^{2} },
\end{equation}
which implies that $\omega_{eff}$ does not depend on the parameter $k$ and hence similarly cannot cross the phantom divide line. Here, $\omega_{eff}$ crosses $-1$ at $Q/Q_{0}\simeq 0.4692$.\\

\begin{figure}[H]
\centering
\begin{subfigure}{0.45\textwidth}
    \includegraphics[width=\textwidth]{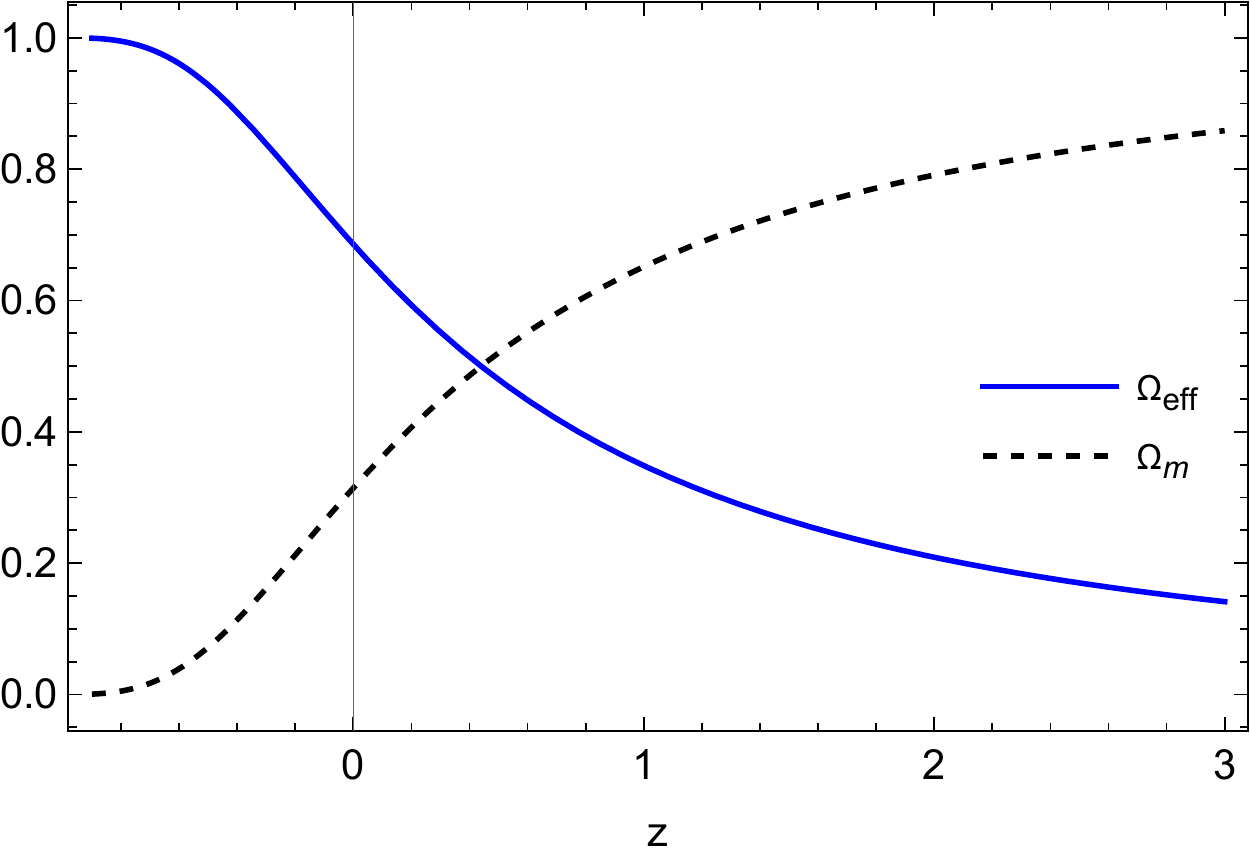}
    \caption{}
    \label{Om log}
\end{subfigure}
\hfill

\caption{Evolution of $\Omega_{eff}$ and $\Omega_{m}$ as functions of redshift $z$ for $k=1$.}
\label{fig:6}
\end{figure}

In fig. \ref{fig:6}, the cosmological evolution of $\Omega_{eff}$ and $\Omega_{m}$ for the parameter $k=1$ is presented. Furthermore, the intersection of $\Omega_{eff}$ and $\Omega_{m}$ is the crossover point $z_{eff}$ at which $\Omega_{eff}=\Omega_{m}$. The universe reaches the matter-dominated stage as $z$ decreases. When $z<z_{eff}$, dark energy eventually overcomes matter. The present values for the parameters are obtained as $\Omega_{eff}^{(0)}=0.68$, $\Omega_{m}^{(0)}=0.314$, and $z_{eff}\approx 0.4$. 

\subsection{Model III}

In this subsection, we consider the combination of exponential and logarithm $F(Q)$ theory to study the cosmological evolution and crossing of phantom divide line. The functional form read as 
\begin{equation}
F(Q)= \beta \left[ Q_{0} \left( \frac{v Q_{0}}{Q} \right) ^{-1/2} ln \left( \frac{v Q_{0}}{Q}\right)-  Q (1- e^{v Q_{0}/Q})\right], 
\end{equation} 
with $\beta= \frac{\Omega_{m}^{0}-1}{2 v^{-1/2} + (1-e^{v}(1-2v))}$, where $v$ is a constant.

%
%

\begin{widetext}

\begin{figure}[]
\centering
\begin{subfigure}{0.45\textwidth}
    \includegraphics[width=\textwidth]{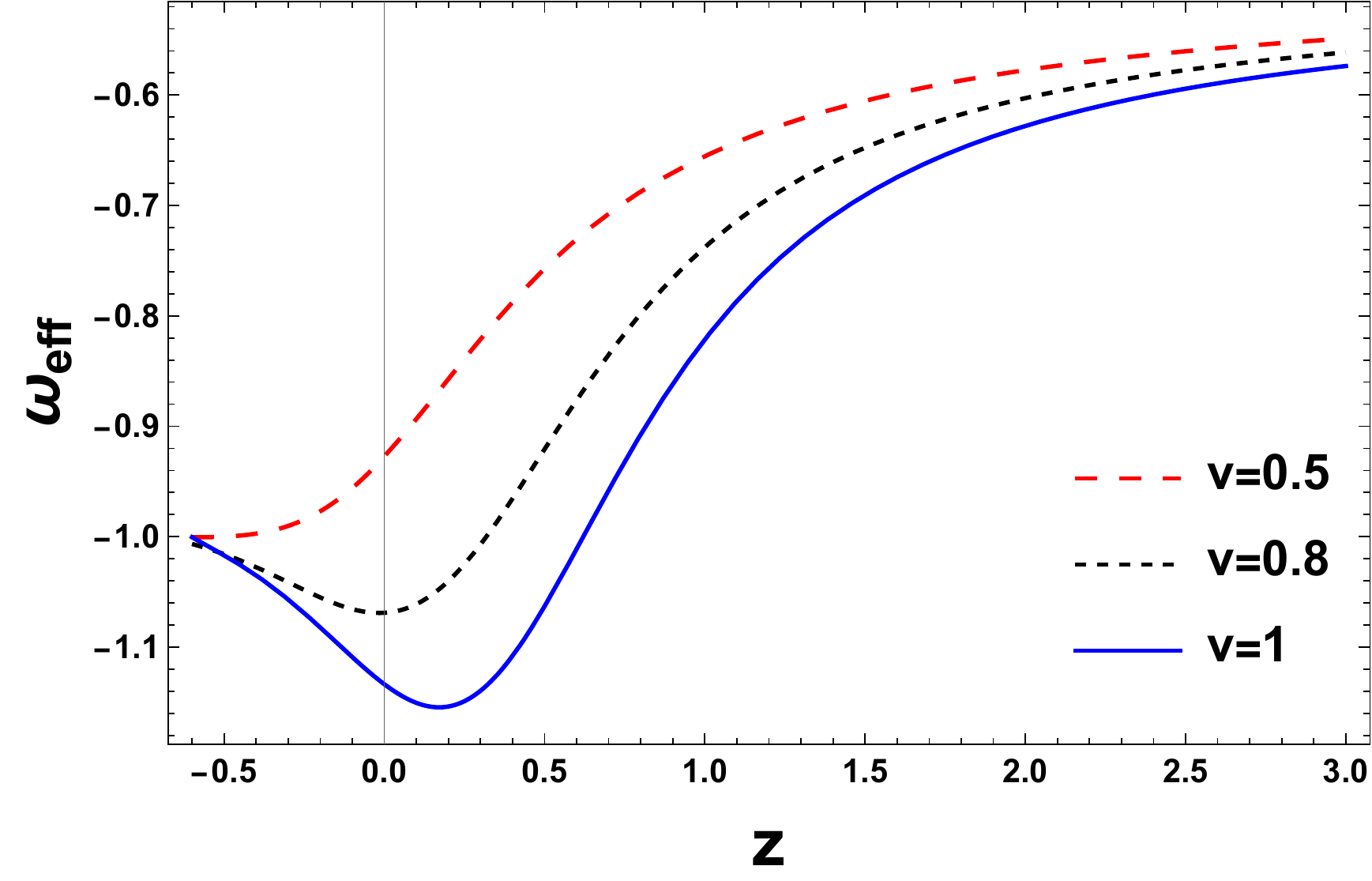}
    \caption{}
    \label{W+combine}
\end{subfigure}
\hfill
\begin{subfigure}{0.45\textwidth}
    \includegraphics[width=\textwidth]{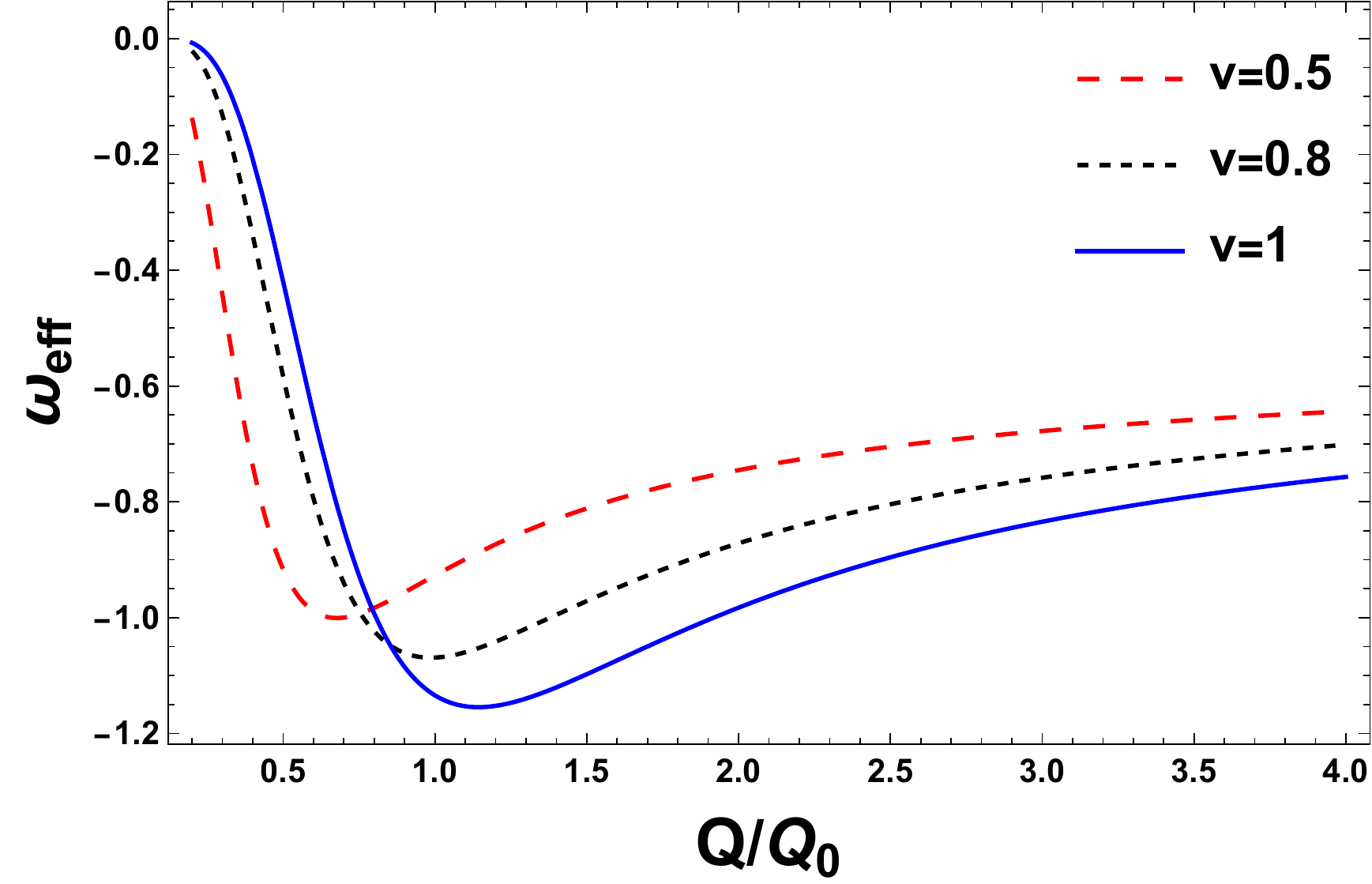}
    \caption{}
    \label{WQ+combine}
\end{subfigure}
\hfill
\begin{subfigure}{0.45\textwidth}
    \includegraphics[width=\textwidth]{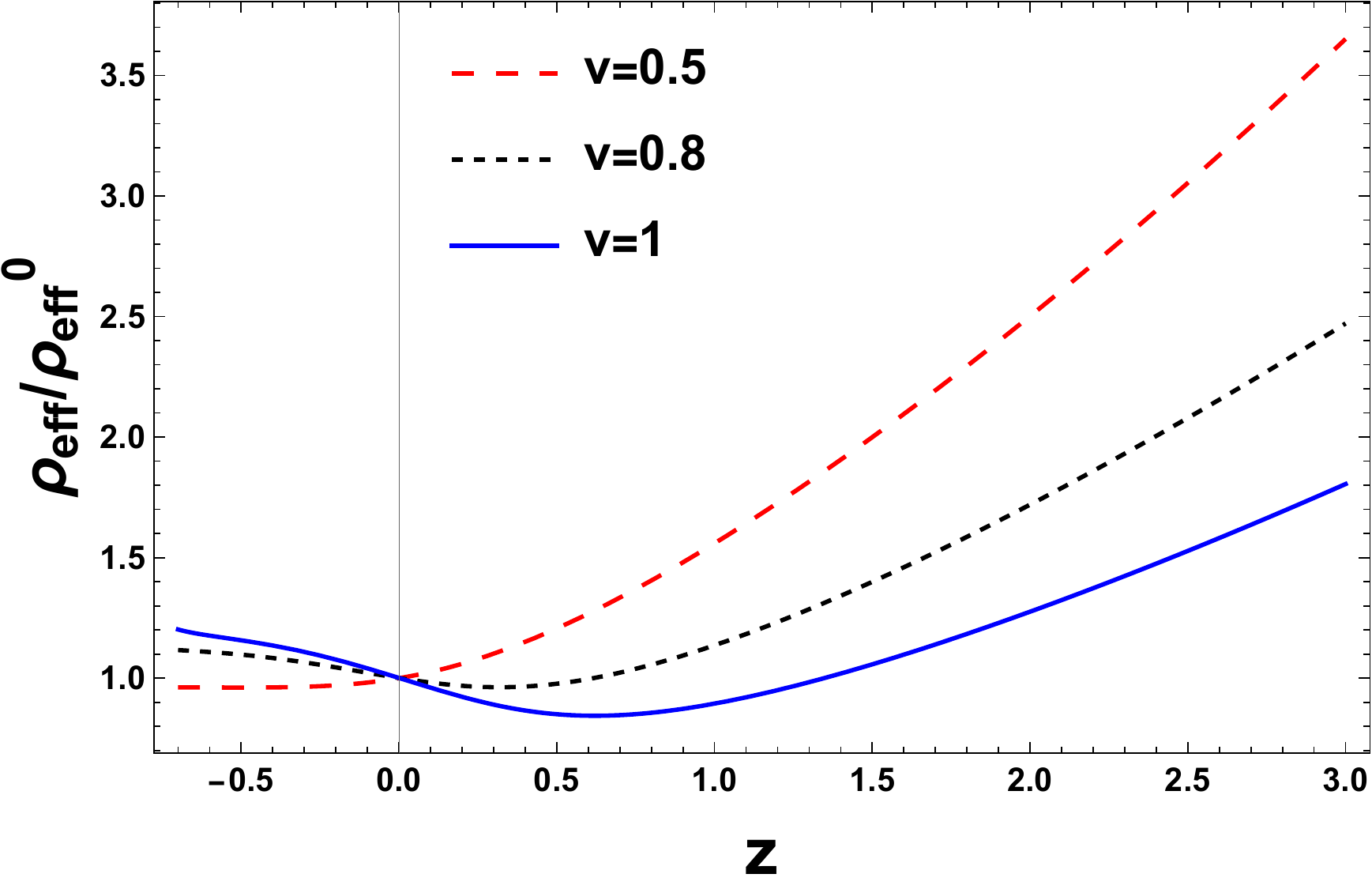}
    \caption{}
    \label{rho-cmbine}
\end{subfigure}
        
\caption{Evolution of $\omega_{eff}$ as a function of redshift $z$ and $Q/Q_{0}$ for $v=0.5$, $0.8$ and $1$. The subplot (c) shows $\rho_{eff}/\rho_{eff}^{0}$ as a function of redshift $z$ for $v=0.5$, $0.8$ and $1$.}
\label{fig:4}
\end{figure}

\end{widetext}

Fig. \ref{W+combine} indicates the behavior of $\omega_{eff}$ versus redshift $z$ for $v=0.5$, $v=0.8$ and $v=1$. We have considered the values of model parameters constrained in ref. \cite{Linder/2010,Bamba/2011}. It is clear that $\omega_{eff}$ shows a transition from a non-phantom phase to a phantom phase with the present values as $-0.927$, $-1.068$, and $-1.133$ \cite{Wu/2011}, respectively. It is noted that the universe crosses the phantom divide line and may approach to $\omega_{eff}=-1$ in the near future.\\
We can also see in fig. \ref{WQ+combine} that $\omega_{eff}$ crosses $-1$ at $Q/Q_{0}= 0.764$ and $0.806$ for $v=0.8$ and $v=1$, respectively.  Furthermore, the evolution of $\rho_{eff}/\rho_{eff}^{0}$ is presented in fig. \ref{rho-cmbine} which shows that effective density becomes dominant in near future.

\textbf{Crossing of the phantom divide line:}\\

Note that from eq. \eqref{14}, $\omega_{eff}$ crosses $-1$ if $4\dot{H} \left( \frac{F_{Q}+2QF_{QQ}}{F-2QF_{Q}} \right)$ changes its sign.

For the investigation, if we concentrate on $\frac{vQ_{0}}{Q} \ll 1$, we obtained the following using approximation to second order:
\begin{equation}
\frac{F}{Q} -2 F_{Q} \approx \beta \sqrt{\frac{Q_{0}v}{Q}}\left[ \frac{2}{v} + \sqrt{\frac{Q_{0}v}{Q}} \left( 1+ \frac{3}{2} \frac{Q_{0}v}{Q} \right) \right] ,
\end{equation}
and 
\begin{equation}
F_{Q}+2QF_{QQ} \approx -\beta \sqrt{\frac{Q_{0}v}{Q}} \left[ \frac{1}{v} - \frac{3}{2} \left(  \frac{Q_{0}v}{Q} \right) ^{3/2} \right] .
\end{equation}
Here, the equations have different sign implies that $\frac{F_{Q}+2QF_{QQ}}{F/Q-2F_{Q}}$ does change its sign for the critical point $\frac{Q_{0}v}{Q}= \left( \frac{2}{3v}\right) ^{2/3}$. Henceforth, $\omega_{eff}$ crosses the phantom divide line in the combined $f(Q)$ model. \\

\begin{figure}[H]
\centering
\begin{subfigure}{0.45\textwidth}
    \includegraphics[width=\textwidth]{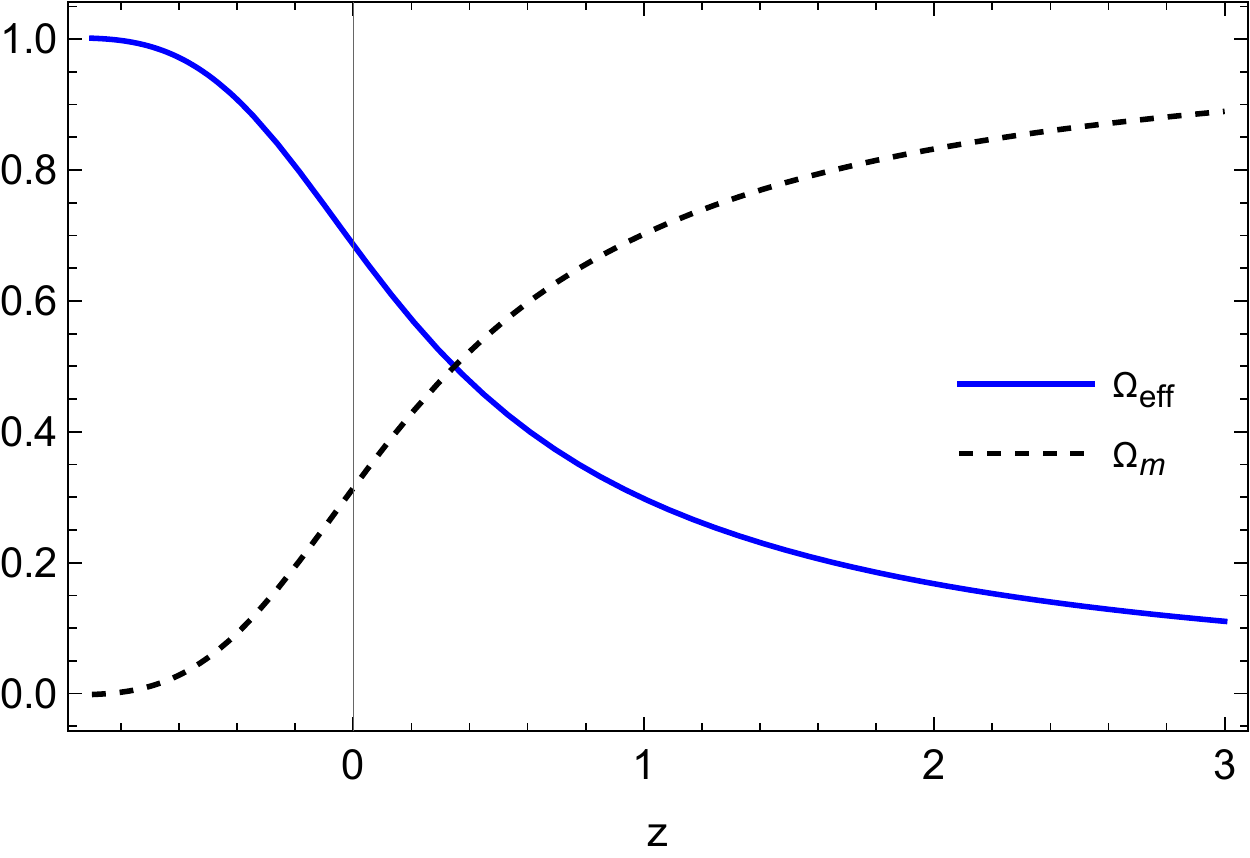}
    \caption{}
    \label{om5 comb}
\end{subfigure}
\hfill
        
\caption{Evolution of $\Omega_{eff}$ and $\Omega_{m}$  as functions of redshift $z$ for $v=0.5$.}
\label{fig:7}
\end{figure}

Figs. \ref{fig:7} shows the evolution of $\Omega_{eff}$ and $\Omega_{m}$ for value of parameter $v=0.5$. The crossover point $z_{eff}$ for this model is $z_{eff}\approx 0.34, 0.27, 0.25$, respectively for $v=0.5$, $v=0.8$, and $v=1$. The universe reaches the matter-dominated stage as $z$ decreases. When $z<z_{eff}$, dark energy eventually overcomes matter. The present values for the parameters are obtained as $\Omega_{eff}^{(0)}\approx 0.68$, $\Omega_{m}^{(0)}\approx 0.314$ for all the three values of $v$. 

\section{Statefinder Diagnostics}  \label{section 5}

A new set of geometrical diagnostics for dark energy with the goal of distinguishing and classifying different dark energy models is proposed by Sahni et al.\cite{Sahni/2003} and Alam et al. \cite{Alam/2003}. The scale factor $a(t)$ and its derivatives up to the third order are used to construct the statefinder diagnostics. Since the statefinder pair $\lbrace r,s \rbrace$ has a geometrical nature, the statefinder parameters are more universal parameters to analyze dark energy models. The diagnostic pair is defined as follows:
\begin{align*}
r &= \frac{\dddot{a}}{a H^3}, \\
s & = \frac{r-1}{3(q-\frac{1}{2})}.
\end{align*}

\begin{widetext}

\begin{figure}[H]
\centering
\begin{subfigure}{0.45\textwidth}
    \includegraphics[width=\textwidth]{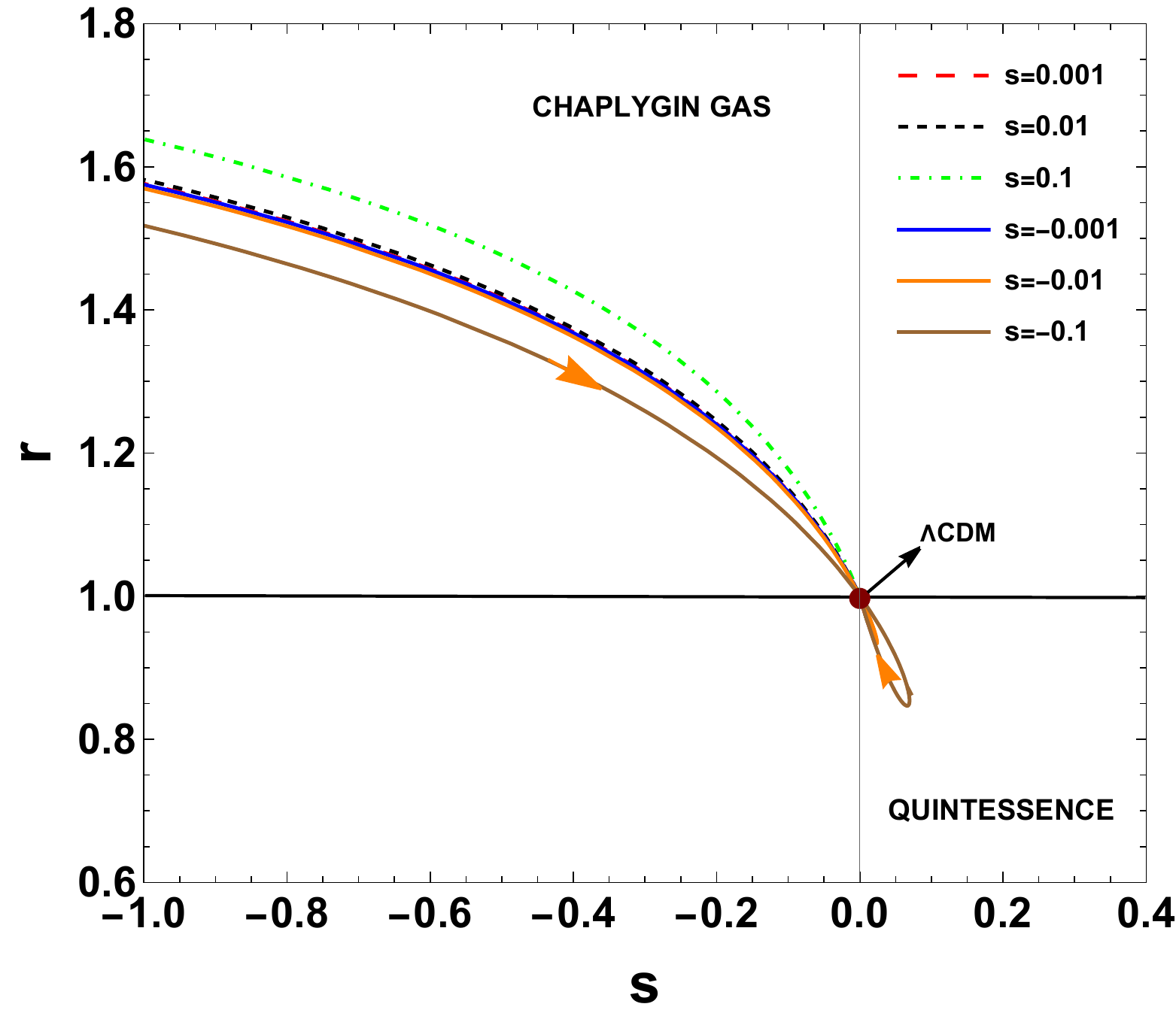}
    \caption{}
    \label{R-S Exp}
\end{subfigure}
\hfill
\begin{subfigure}{0.45\textwidth}
    \includegraphics[width=\textwidth]{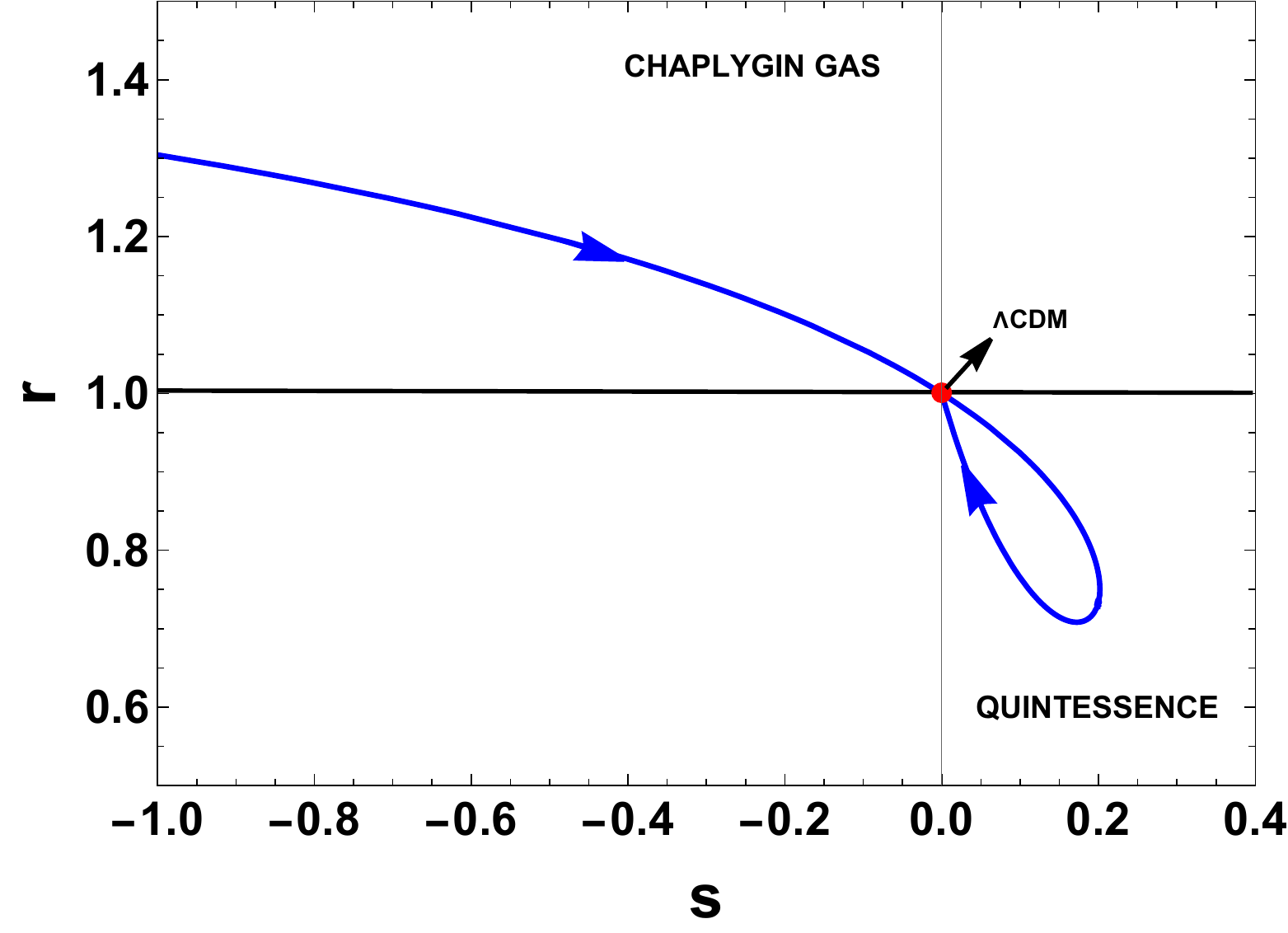}
    \caption{}
    \label{R-S log}
\end{subfigure}
\hfill
\begin{subfigure}{0.45\textwidth}
    \includegraphics[width=\textwidth]{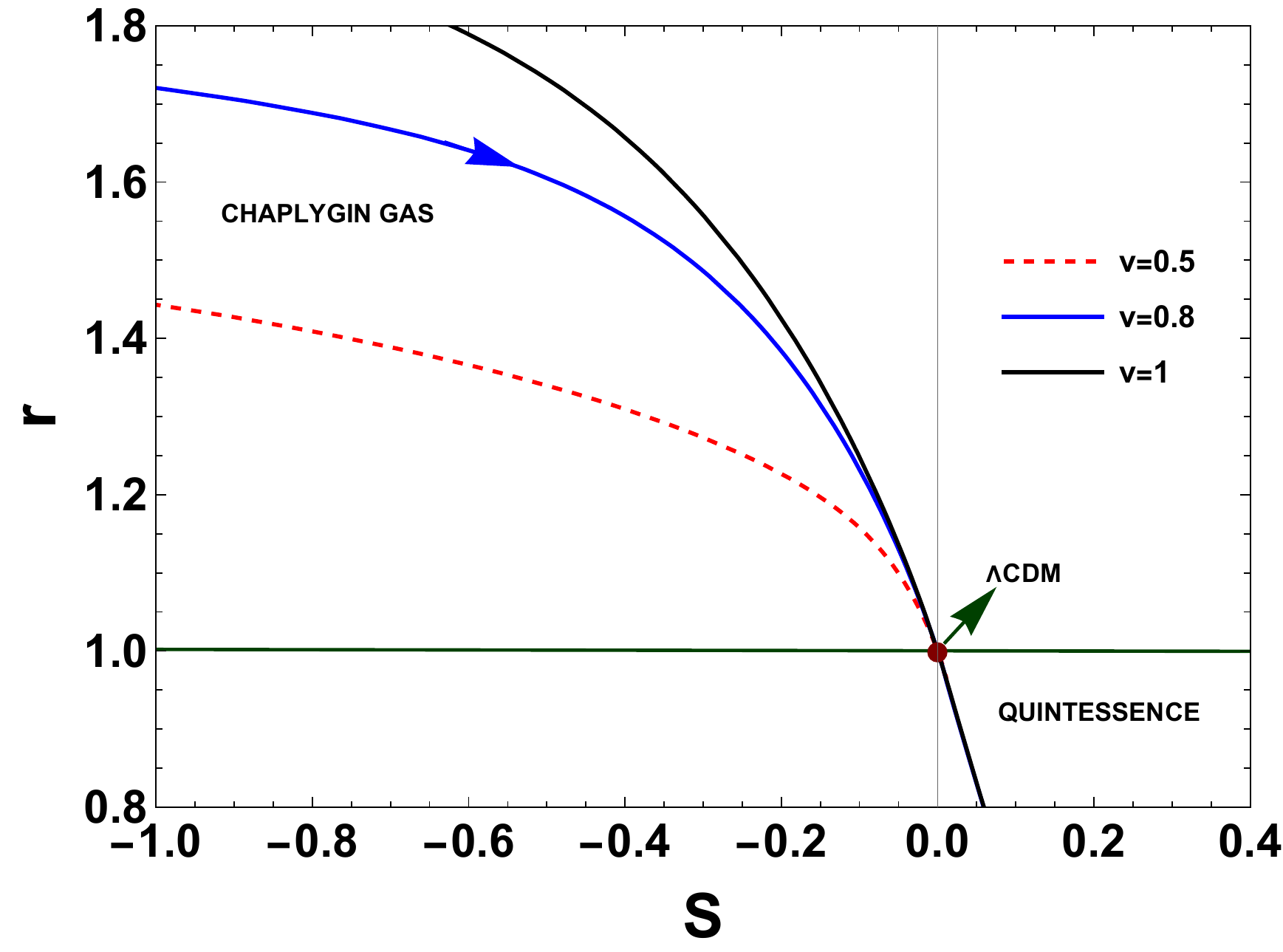}
    \caption{}
    \label{R-S Comb}
\end{subfigure}
        
\caption{Trajectories of $s-r$ plane for model I, model II and model II, respectively.}
\label{fig:8}
\end{figure}

\end{widetext}

Here, $q=-\frac{a \ddot{a}}{\dot{a}^2}$. For more on statefinder analysis, one can check References \cite{Raja/2021,Pacif/2021,Shabani/2017}. We plot the trajectories in $\lbrace s,r \rbrace$ and $\lbrace q,r \rbrace$ place for all the three models. It is clear from the figs. \ref{R-S Exp}, \ref{R-S log}, and \ref{R-S Comb} that the dark energy models always approaches to $\Lambda$CDM model, i.e. $(s,r)=(0,1)$ in the late time evolution. It is interesting to note that the trajectory in the case of model III passes through the $\Lambda$CDM point but may not approach to the $\Lambda$CDM at late times.

\begin{widetext}

\begin{figure}[]
\centering
\begin{subfigure}{0.45\textwidth}
    \includegraphics[width=\textwidth]{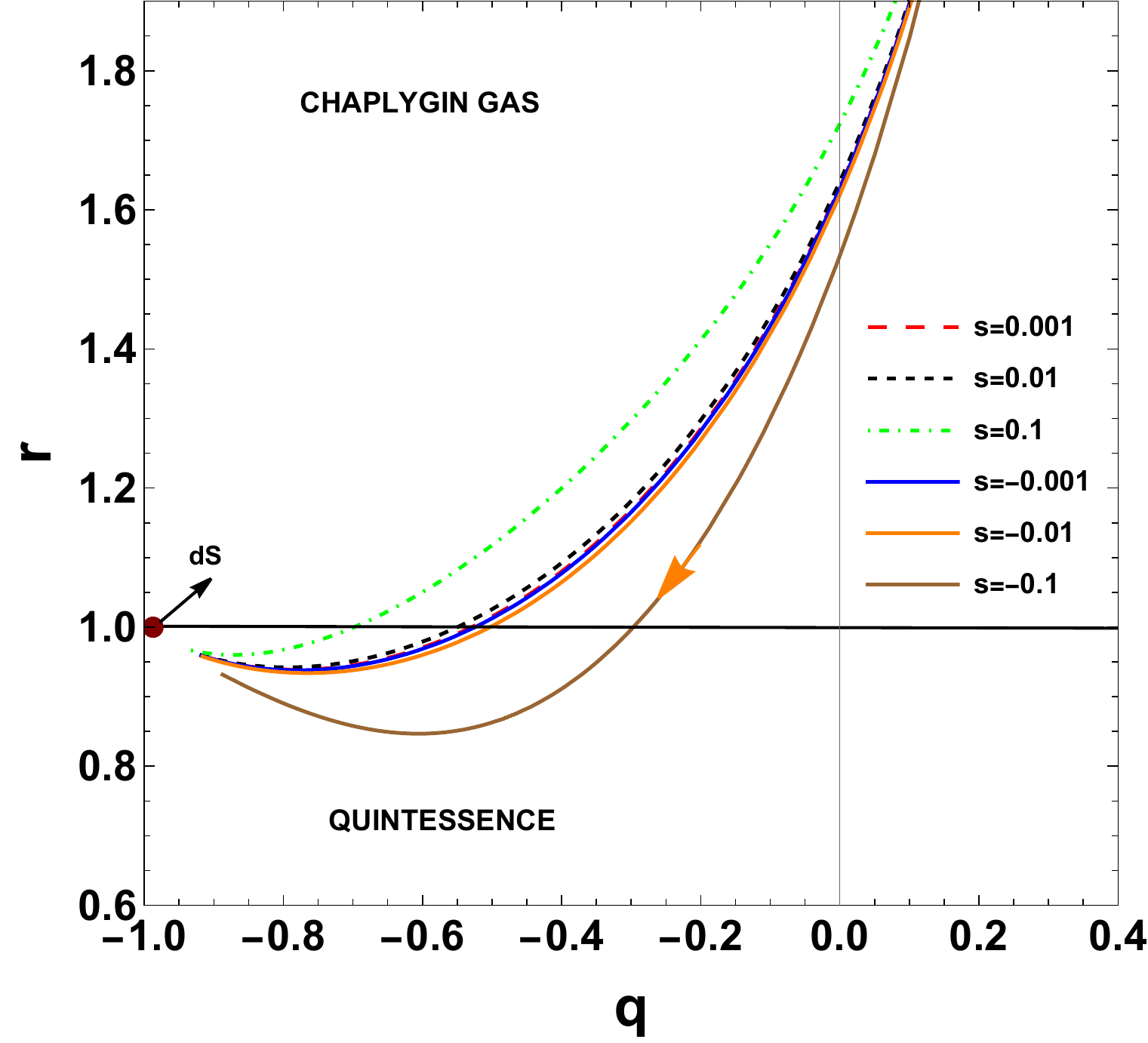}
    \caption{}
    \label{Q-R Exp}
\end{subfigure}
\hfill
\begin{subfigure}{0.45\textwidth}
    \includegraphics[width=\textwidth]{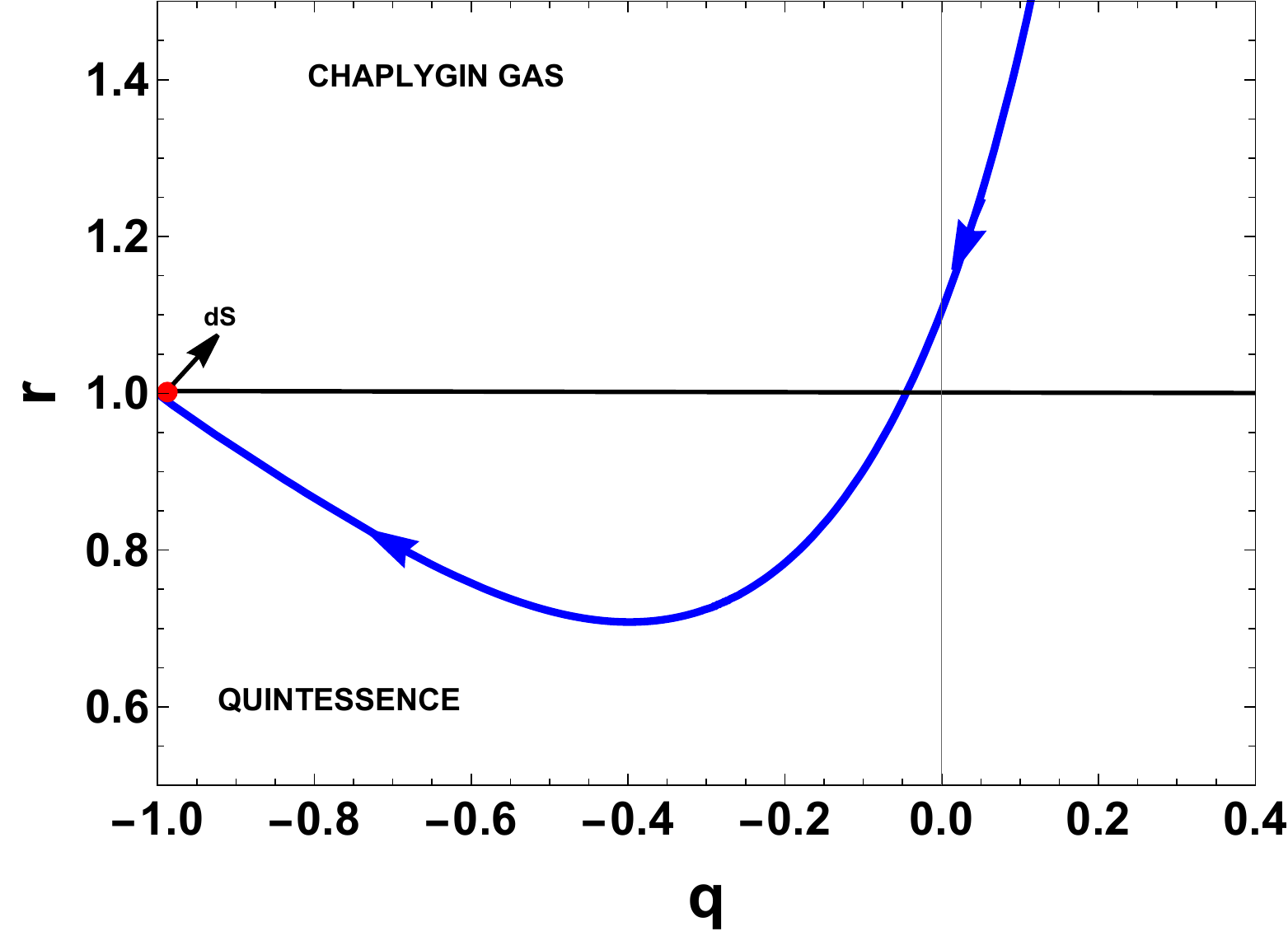}
    \caption{}
    \label{Q-R log}
\end{subfigure}
\hfill
\begin{subfigure}{0.45\textwidth}
    \includegraphics[width=\textwidth]{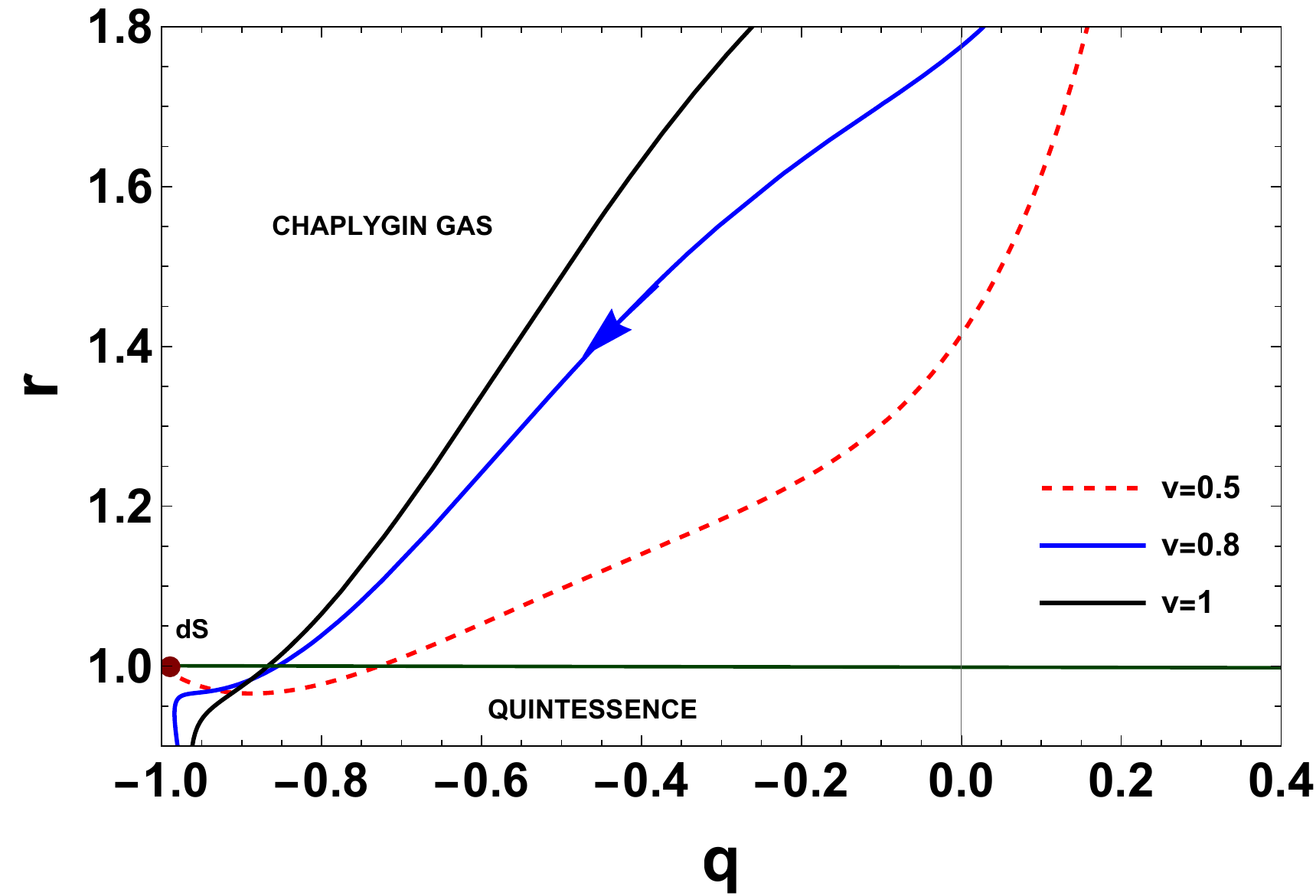}
    \caption{}
    \label{Q-R Comb}
\end{subfigure}
        
\caption{Trajectories of $q-r$ plane for model I, model II and model II, respectively.}
\label{fig:9}
\end{figure}

\end{widetext}

Fig. \ref{Q-R Exp}, \ref{Q-R log}, and \ref{Q-R Comb} plot the trajectories in $q-r$ plane for model I, model II and Model III. The dS model (steady state) have been shown by the point $(q,r)=(-1,1)$. The recent phase transition is satisfactorily explained by the signature flip of the deceleration parameter $q$ from positive to negative. One can observe that, for different values of respective parameters for model I and model II, the dark energy model approaches the $dS$ point as Chaplygin gas, $\Lambda$CDM and quintessence at late time evolution. Furthermore, in case of model III, the trajectories deviates from $dS$ model at late times.

\section{Linear Perturbations Analysis} \label{section 6}

In the present section, we investigate the stability of the cosmological models considered under the homogeneous linear perturbations. One can define the first order perturbation for the Hubble and density parameter as \cite{Farrugia/2016,Dombriz/2012,Anagnost/2021}
\begin{align} 
\widetilde{H(t)} & = H(t)(1+\delta)\\
\widetilde{\rho(t)} & = \rho(t)(1+\delta_{m}).  
\end{align}

\begin{widetext}

\begin{figure}[H]
\centering
\begin{subfigure}{0.45\textwidth}
    \includegraphics[width=\textwidth]{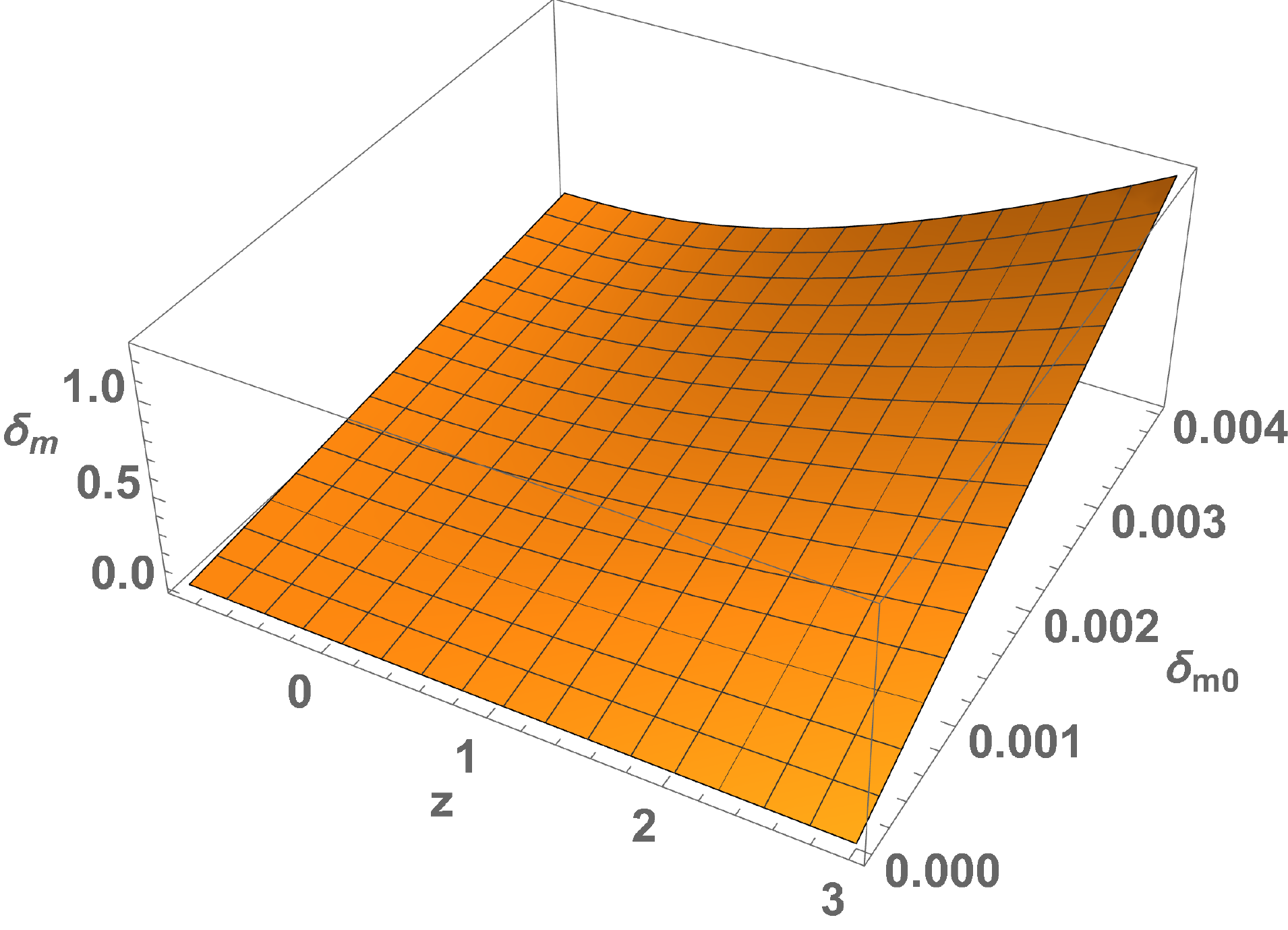}
    \caption{}
    \label{dm exp}
\end{subfigure}
\hfill
\begin{subfigure}{0.45\textwidth}
    \includegraphics[width=\textwidth]{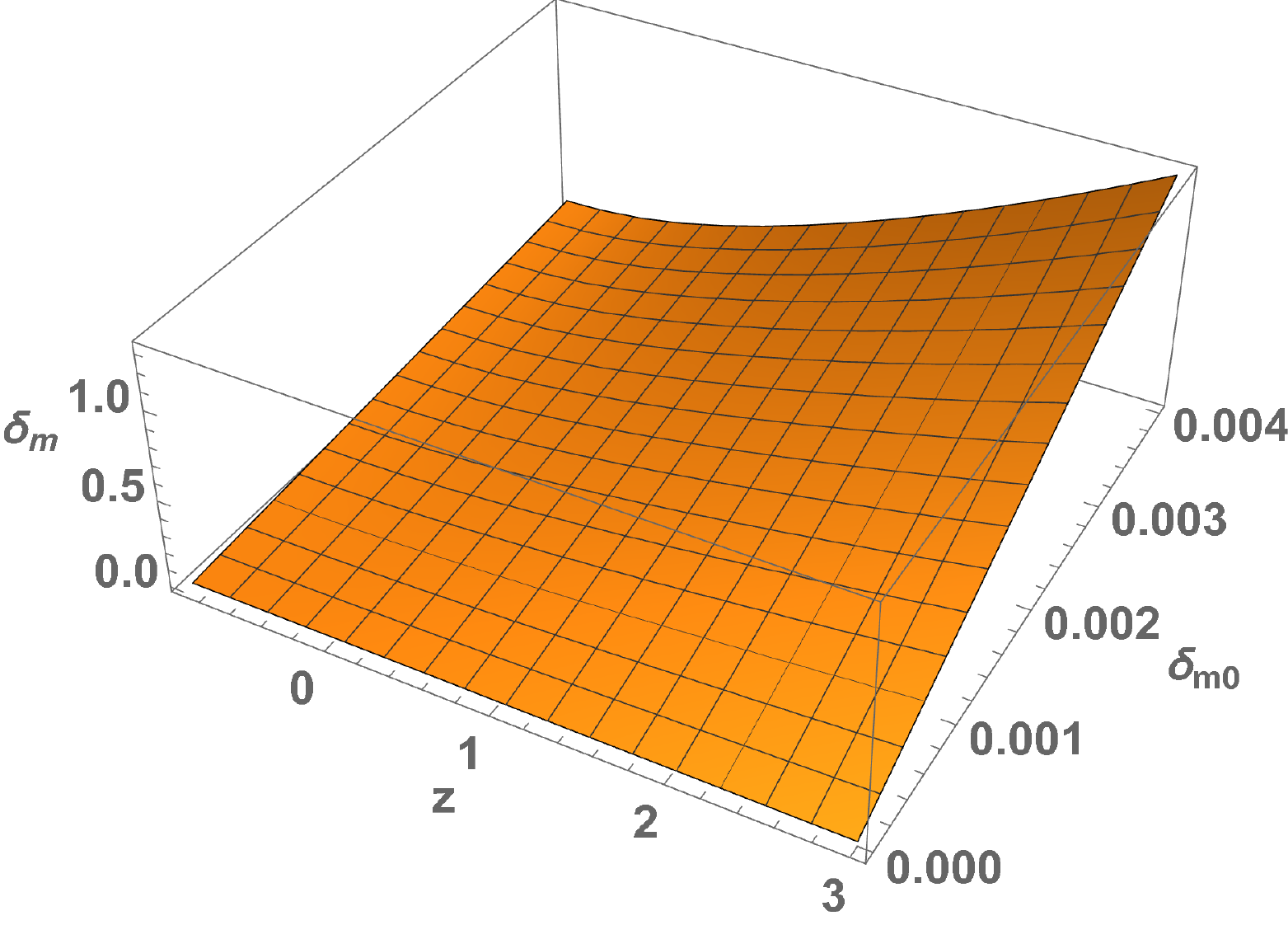}
    \caption{}
    \label{dm log}
\end{subfigure}
\hfill
\begin{subfigure}{0.45\textwidth}
    \includegraphics[width=\textwidth]{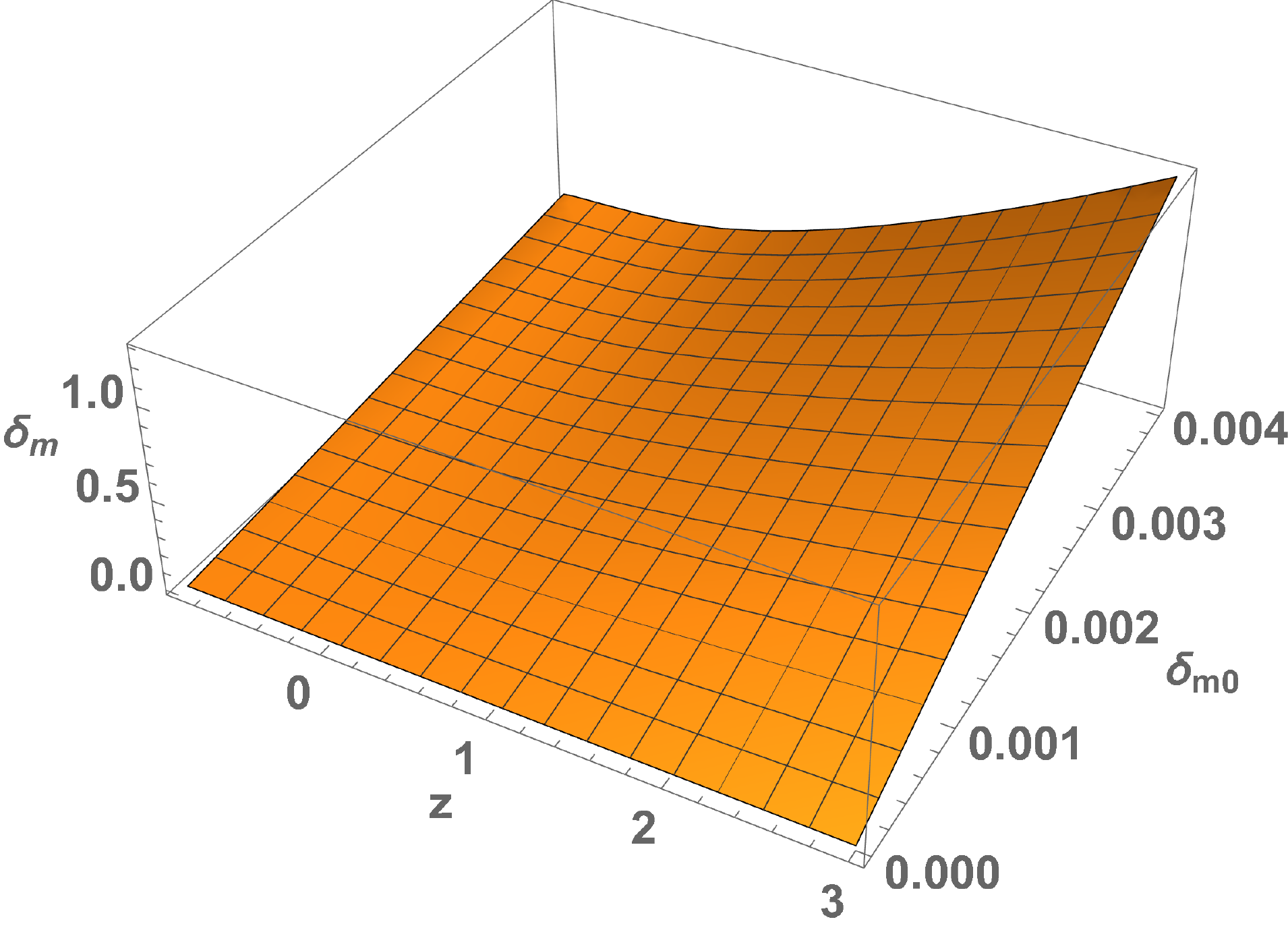}
    \caption{}
    \label{dm-Comb}
\end{subfigure}
        
\caption{Evolution of $\delta_{m}$ for model I, model II and model III with $0<\delta_{m_{0}}<0.004$, respectively.}
\label{fig:10}
\end{figure}

\end{widetext}

Here, $\widetilde{H(t)}$, and $\widetilde{\rho(t)}$ represents the perturbed Hubble and density parameter, $\delta$ and $\delta_{m}$ are the perturbation terms, respectively. Now eqns. \eqref{8} and \eqref{9} can be simplified to 
\begin{align} \label{Hd}
\frac{f}{2} - Q f_{Q} & = \rho\\ 
\left(f_{Q}+2Qf_{QQ}\right)\dot{H} & = \frac{1}{2}\left(\rho + p\right). \label{Hdd}
\end{align}
which are satisfied by the perturbed quantities with the continuity equation. We can obtain the perturbed $f$ and $f_{Q}$ as $\delta f= f_{Q} \delta Q$ and $\delta f_{Q}= f_{QQ} \delta Q$ with $\delta Q= 12 H \delta H$. Now, using the continuity equation and \eqref{Hd}, we get the following equations: 

\begin{align}
Q \left(f_{Q}+2 Q f_{QQ}\right)\delta &= -\rho \delta_{m},\\
\dot{\delta_{m}} + 3 H (1+\omega) \delta & =0.
\end{align}
Solving the above equations for $\delta$ and $\delta_{m}$, we have
\begin{equation}
\dot{\delta_{m}}- \frac{3 H(1+\omega)\rho}{Q(f_{Q}+2 Qf_{QQ})} \delta_{m}=0.
\end{equation}
Simplifying the above equation using eqn. \eqref{Hdd}, the solution read as 
\begin{align}
\delta_{m} & = \delta_{m_{0}} H\\
\delta & = \delta_{0}\frac{\dot{H}}{H}= \delta_{0}(1+z)\frac{dH}{dz}, 
\end{align}
where $\delta_{m_{0}}$ is constant. Also $\delta_{0}= -\frac{\delta_{m_{0}}}{3(1+\omega)}$.
We present the behavior of perturbation terms $\delta_{m}$ and $\delta$ with respect to redshift $z$ in figs. \ref{dm exp}, \ref{dm log}, and \ref{dm-Comb} for model I, model II and model III, respectively. It is clear that $\delta_{m}$ and $\delta$ decay rapidly at late times and depicts the stability of our considered models. 

\begin{figure}[H]
\centering
\begin{subfigure}{0.45\textwidth}
    \includegraphics[width=\textwidth]{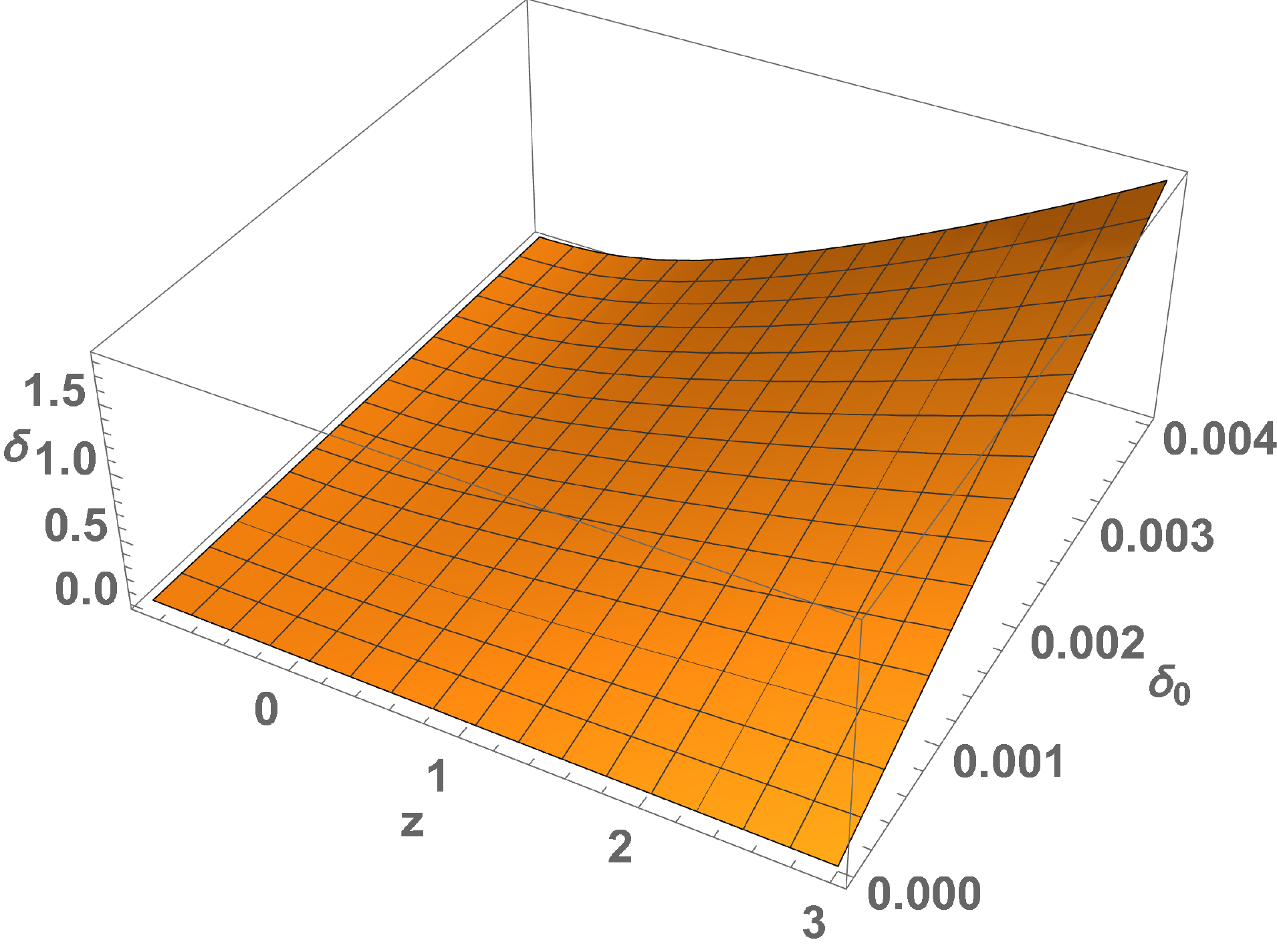}
    \caption{}
    \label{d exp}
\end{subfigure}
\hfill
\begin{subfigure}{0.45\textwidth}
    \includegraphics[width=\textwidth]{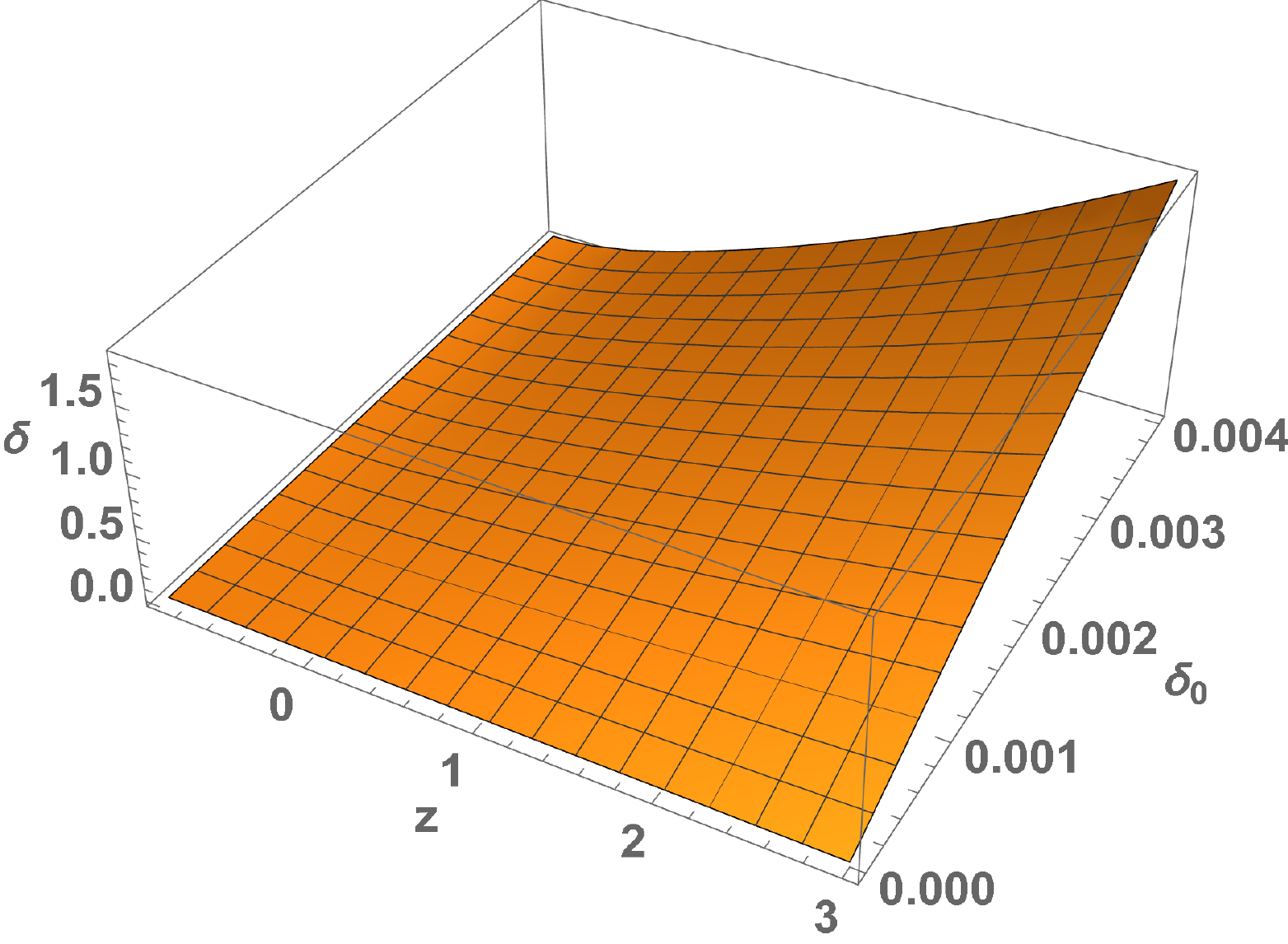}
    \caption{}
    \label{d log}
\end{subfigure}
\hfill
\begin{subfigure}{0.45\textwidth}
    \includegraphics[width=\textwidth]{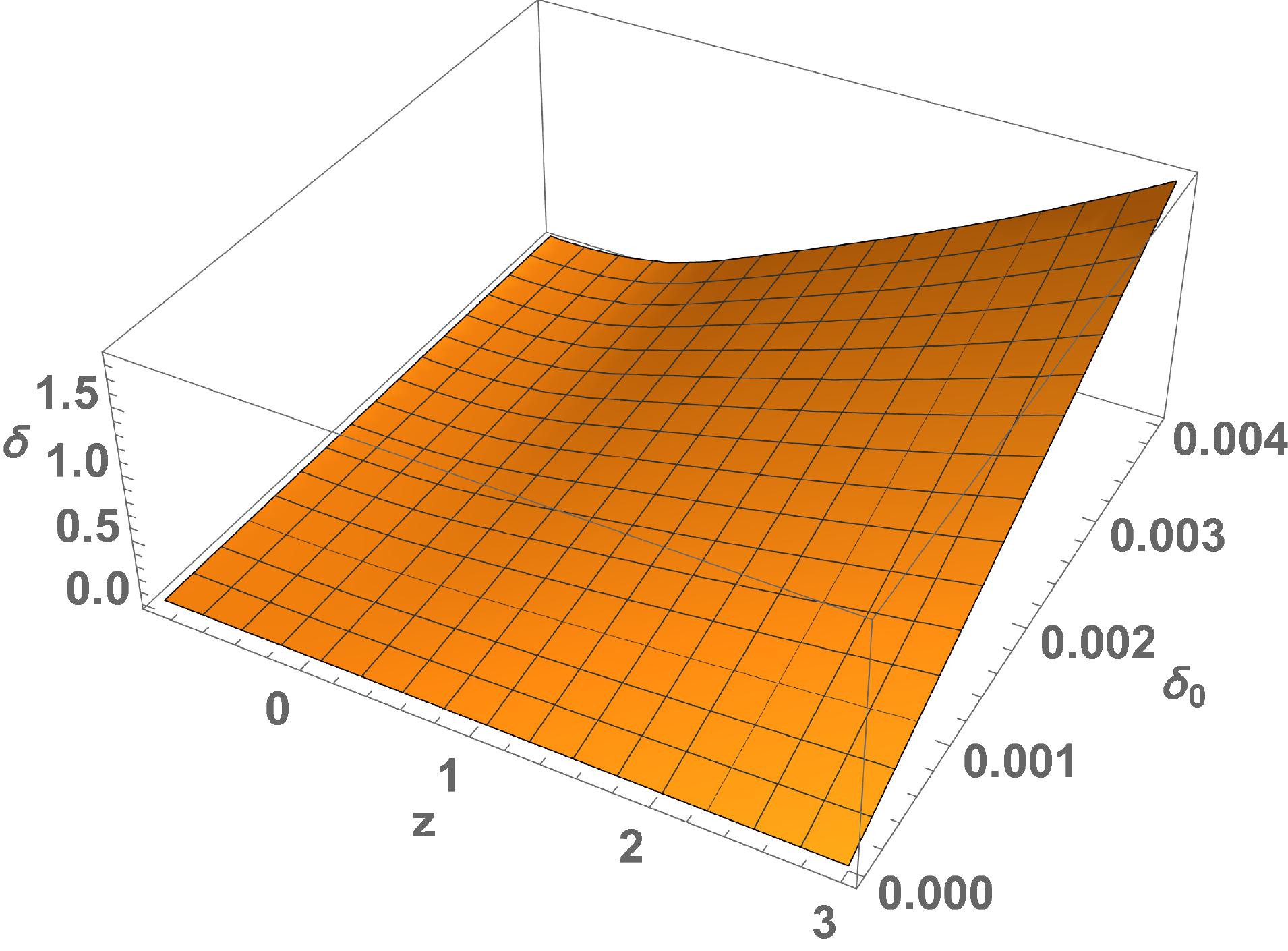}
    \caption{}
    \label{d-Comb}
\end{subfigure}
        
\caption{Evolution of $\delta$ for model I, model II and model III with $0<\delta_{0}<0.004$, respectively.}
\label{fig:11}
\end{figure}

It is seen that the evolution of $\delta_{m}$ is similar for all the model parameters. The evolution of $\delta$ is presented in figs. \ref{d exp}, \ref{d log}, and \ref{d-Comb}  for $s=0.1$, $k=1$, and $v=0.8$. 

\section{Reconstruction of F(Q)}\label{section 7}

The cosmographic technique allows to explore the dynamics of the universe using kinematic quantities that are independent of the background cosmology. Thus, the late-time expansion history of the universe can be examined to the cosmological principle in order to obtain information on the dark energy attributes and the nature of gravity. Pad'e approximations \cite{Gruber/2014} are one of the most dependable cosmographic strategies for ensuring steady behavior at high redshifts. The corresponding Hubble expansion is obtained as \cite{Capo/2022}
\begin{equation}
H(z)= H_{0} \frac{M(z; q_{0}, j_{0})}{N(z; q_{0},j_{0})},
\end{equation}
where 
\begin{align*}
M(z;q_{0},j_{0}) & = 2(1 + z)^{2} (3+z +j_{0}z-3 q_{0}^2 z-q_{0}(3+z))^{2}\\
N(z;q_{0},j_{0}) & = 18 + 6(5+2j_{0})z+(14+7 j_{0}+2 j_{0}^2) z^2 + 9 q_{0}^4 z^2 \\
                 & + 18 q_{0}^3 z(1+z)- 2 q_{0} (6+5 z)(3+(4+j_{0})z) \\
                 & +q_{}^2 (18+30 z +(17-9 j_{0})z^2 ).
\end{align*}

We use the values of cosmographic parameters as $H_{0}=69$, $q_{0}= -0.73$ and $j_{0}=2.84$ \cite{Capo/2020}. The term $F_{Q}$ can be written in terms of Hubble parameter as a function of redshift $z$, i.e. $F_{Q}=\frac{F'(z)}{12 H(z)H'(z)}$, where $F'(z)$ is the derivative of $F$ with respect to $z$. Hence, we can write eqn. \eqref{8} as 
\begin{equation}
\frac{H(z) F'(z)}{H'(z)}- F(z)+6 H_{0}^2 \Omega_{m_{0}} (1+z)^3 - 6 H^2=0.
\end{equation}

\begin{figure}[H]
\centering
    \includegraphics[width=0.45\textwidth]{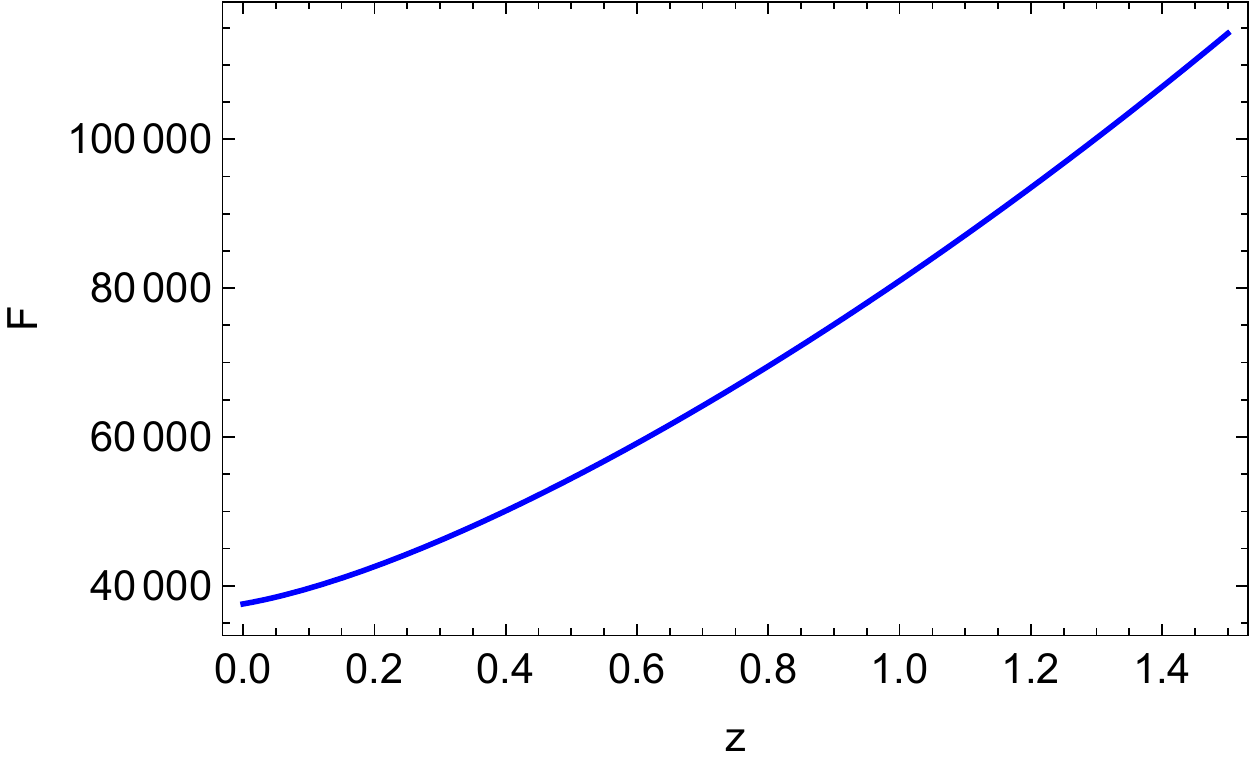}
\caption{Evolution of $F$ versus $z$.}
\label{fig:12}
\end{figure}

According to the analysis in \cite{Capo/2022}, we use the initial condition $F_{Q}(z=0)=1$ leading to $F_{0}= 6 H_{0}^2 (1+\Omega_{m_{0}})$. Solving numerically gives the similar trajectory in fig. \ref{fig:12} mentioned in \cite{Capo/2022} as 
$F(Q)= \zeta +\sigma Q^n$. 

To see if the phantom divide line crosses for $F(Q)= \zeta +\sigma Q^n$, we look at whether $4 \dot{H} \left(\frac{F_{Q}+2Q F_{QQ}}{F-2 Q F_{Q}}\right)$ changes its sign. 
\begin{align*}
\frac{F}{Q}- 2 F_{Q} & = \frac{ \zeta + \sigma Q^n (1-2n)  }{Q}\\
F_{Q}+2QF_{QQ} & = -n (1-2n) \sigma  Q^{n-1}.
\end{align*}

According to the trajectory of $F$, we can assure that $\zeta>0$ and $\sigma>0$. Furthermore, the equations above have different signs for any value of $n$. This depicts that $\omega_{eff}$ may cross the phantom divide line in the obtained $F(Q)$ model.

\section{Conclusion}
\label{section 8}

The newly proposed extended symmetric teleparallel or the $f(Q)$ gravity theory is an exciting extension of Einstein's new general relativity, employing a non-metricity $Q$ rather than curvature. Recently, there has been renewed interest in Einstein's new equivalent theories for obtaining cosmic acceleration. We may discover that Einstein predicted the cosmic acceleration not necessarily through the cosmological constant but also through the equivalent representation of general relativity.\\
We investigated the cosmic evolution in the exponential and logarithmic $F(Q)$ theories. In the exponential $F(Q)$ model, the universe is observed to be in the non-phantom regime without crossing the phantom divide line. Similarly, the logarithmic $F(Q)$ theory with one model parameter does not permit phantom divide crossing.\\
Furthermore, the crossing of the phantom division line from a non-phantom phase to a phantom phase is a remarkable characteristic of the combination of exponential and logarithmic $F(Q)$ model. It is also  worth noting that the phantom divide crossing is compatible with the observations. In section \ref{section 5}, we discussed the statefinder diagnostics to differentiate various dark energy models. It is observed that the dark energy models always approaches to $\Lambda$CDM model, i.e. $(s,r)=(0.1)$ in the late time evolution.  Also, it is noted that the trajectory in the case of model III passes through the $\Lambda$CDM point but may not approach to the $\Lambda$CDM at late times. There is a large deviation  from $\Lambda$CDM point in case of the combined $F(Q)$ model. \\
Further, in section \ref{section 6}, the perturbed equations are solved analytically and depends on the Hubble parameter. It is either decaying to zero at late times or being identically zero for the Hubble perturbation. The matter perturbation, on the other hand, likewise decays to zero. As a result, this solution can be regarded as stable with respect to small perturbations and hence implying the stability of $F(Q)$ models. We also used the numerical inversion approach to rebuild the $F(Q)$ using the relation $Q=6 H^2$. We discovered that the function $F(Q)= \zeta+\sigma Q^n$ provides the best analytical fit and may cross the phantom divide line at late times. As a result, the work demonstrates that the non-metric theories can be used to drive the cosmic acceleration.

\section*{Acknowledgements}

SA acknowledges CSIR, New Delhi, India for JRF. PKS acknowledges CSIR, New Delhi, India for financial support to carry out the Research project [No.03(1454)/19/EMR-II Dt.02/08/2019].


\begin{thebibliography}{90}

\bibitem{Riess/1998} A.G. Riess et al., Astron. J., \textbf{116}, 1009 (1998).

\bibitem{Perlmutter/1999}  S. Perlmutter et al., Astrophys. J., \textbf{517}, 565 (1999).

\bibitem{Ade/2016} P.A.R. Ade et al., Astron. Astrophys., \textbf{594}, A13 (2016).

\bibitem{Aghanim/2020} N. Aghanim et al., A\& A, \textbf{641}, A6 (2020).

\bibitem{Buchdahl/1970} H.A. Buchdahl, Month. Not. R. Astron. Soc., \textbf{150}, 1 (1970).

\bibitem{Capo/2008} S. Capozziello, V. F. Cardone, V. Salzano, Phys. Rev. D, \textbf{78}, 063504 (2008).

\bibitem{Cruz/2006} A. de la Cruz-Dombriz, A. Dobado, Phys. Rev. D, \textbf{74}, 087501 (2006).

\bibitem{Capo/2011} S. Capozziello et al., Phys. Rev. D, \textbf{84}, 043527 (2011).

\bibitem{Liu/2012} Di Liu, M. J. Reboucas, Phys. Rev. D, \textbf{86}, 083515 (2012).

\bibitem{Iorio/2012} L. Iorio, E. N. Saridakis, Month. Not. R. Astron. Soc., \textbf{427}, 1555 (2012).

\bibitem{Wang/2020} Deng Wang, David Mota, Phys. Rev. D, \textbf{102}, 063530 (2020).

\bibitem{Nunes/2016} R. C. Nunes, S. Pan, E. N. Saridakis, JCAP, \textbf{08}, 011 (2016).

\bibitem{Nester/1999} J.M. Nester, H.-J. Yo, Chin. J. Phys., \textbf{37}, 113 (1999).

\bibitem{Lazkoz/2019} R. Lazkoz et al., Phys. Rev. D, \textbf{100}, 104027 (2019).

\bibitem{Jimenez/2018} J. Beltran Jimenez, L. Heisenberg, T. Koivisto, Phys. Rev. D, \textbf{98}, 044048 (2018).

\bibitem{Jimenez/2020} J. Beltran Jimenez et al., Phys. Rev. D, \textbf{101}, 103507 (2020).

\bibitem{Mandal/2020} S. Mandal, D. Wang, P.K. Sahoo, Phys. Rev. D, \textbf{102}, 124029 (2020).

\bibitem{Mandal/2020a} S. Mandal, P.K. Sahoo, J.R.L. Santos, Phys. Rev. D, \textbf{102}, 024057 (2020).

\bibitem{Ambrosio/2022} F. D'Ambrosio et al., Phys. Rev. D, \textbf{105}, 024042 (2022).

\bibitem{Bajardi/2020} F. Bajardi, D. Vernieri, S. Capozziello, Eur. Phys. J. Plus, \textbf{135}, 912 (2020).

\bibitem{Mandal/2021} S. Mandal et al., Eur. Phys. J. Plus, \textbf{136}, 760 (2021).

\bibitem{Ayuso/2021} I. Ayuso, R. Lazkoz, V. Salzano, Phys. Rev. D, \textbf{103}, 063505 (2021).

\bibitem{Khyllep/2021} W. Khyllep, A. Paliathanasis, Jibitesh Dutta, Phys. Rev. D, \textbf{103}, 103521  (2021).

\bibitem{Albuquerque/2022} I.S. Albuquerque, N. Frusciante, Phys. Dark Univ., \textbf{35}, 100980 (2022).

\bibitem{Esposito/2022} F. Esposito et al., Phys. Rev. D, \textbf{105}, 084061 (2022).

\bibitem{Atayde/2021} L. Atayde, N. Frusciante, Phys. Rev. D, \textbf{104}, 064052 (2021).

\bibitem{Zhao/2022} D. Zhao, Eur. Phys. J. C, \textbf{82}, 303 (2022).

\bibitem{Frusciante/2021} N. Frusciante,  Phys. Rev. D, \textbf{103}, 044021 (2021).

\bibitem{Dimakis/2021} N. Dimakis, A. Paliathanasis, T. Christodoulakis, Class. Quantum Gravit., \textbf{38}, 225003 (2021).

\bibitem{Barros/2020} B.J. Barros et al., Phys. Dark Univ., \textbf{30}, 100616 (2020).

\bibitem{Harko/2018}  T. Harko et al., Phys. Rev. D, \textbf{98}, 084043 (2018).

\bibitem{Bajardi/2020} F. Bajardi, D. Vernieri, S. Capozziello, Eur. Phys. J. C, \textbf{135}, 912 (2020).


\bibitem{Apos/2006} P. S. Apostolopoulos, N. Tetradis, Phys. Rev. D, \textbf{74}, 064021  (2006).

\bibitem{Wu/2011} P. Wu, H. Yu, Eur. Phys. J. C, \textbf{71}, 1552 (2011).

\bibitem{Nozari/2008} K. Nozari, M. Pourghasemi, JCAP, \textbf{10}, 044 (2008).

\bibitem{Bamba/2010} K. Bamba, C.-Q. Geng, C.-C. Lee, JCAP, \textbf{11}, 001 (2010).


\bibitem{Zhao/2019} J. Lu, X. Zhao, G. Chee, Eur. Phys. J. C, \textbf{79}, 530 (2019).

\bibitem{Karimzadeh/2019} S. Karimzadeh, R. Shojaee, Adv. High Energy Phys., \textbf{2019}, 4026856 (2019).

\bibitem{Yang/2010} L. Yang et al., Phys. Rev. D, \textbf{82}, 103515 (2010).

\bibitem{Bamba/2011} K. Bamba et al., JCAP, \textbf{01}, 021 (2011).

\bibitem{Alam/2003} U. Alam et al., Mon. Not. Roy. Astron. Soc., \textbf{344}, 1057-1074 (2003).


\bibitem{Gorini/2003} V. Gorini, A. Kamenshchik, U. Moschella, Phys. Rev. D, \textbf{67}, 063509 (2003).

\bibitem{Sahni/2003} V. Sahni et al.,  Jetp Lett., \textbf{77}, 201-206 (2003).

\bibitem{Zhang/2005} X. Zhang, Phys. Lett. B, \textbf{611}, 1 (2005). 

\bibitem{Zhang/2008} J. Zhang, X. Zhang, H. Liu, Phys. Lett. B, \textbf{659}, 26-33 (2008).

\bibitem{Setare/2007} M.R. Setare, Phys. Lett. B, \textbf{653}, 116-121 (2007).   

\bibitem{Chang/2007} B. Chang et al., JCAP, \textbf{01}, 016 (2007).

\bibitem{Linder/2010} Eric V. Linder, Phys. Rev. D, \textbf{81}, 127301 (2010). 

\bibitem{Sanjay/2021} S. Mandal, P.K. Sahoo, Phys. Lett. B, \textbf{823}, 136786 (2021).


\bibitem{Hu/2007} W. Hu, I. Sawicki, Phys. Rev. D, \textbf{76}, 064004 (2007).


 \bibitem{Raja/2021} R. Solanki et al., Phys. Dark Univ., \textbf{32}, 100820 (2021).
 
 \bibitem{Pacif/2021} S.K.J. Pacif, S. Arora, P.K. Sahoo, Phys. Dark Univ., \textbf{32}, 100804 (2021).
 
 \bibitem{Shabani/2017} H. Shabani, Int. J. Mod. Phys. D, \textbf{26}, 1750120 (2017).
 
\bibitem{Farrugia/2016}  G. Farrugia, J. L. Said, Phys. Rev. D, \textbf{94}, 124054 (2016).

\bibitem{Dombriz/2012} A. de la C-Dombriz, D. S-Gomez, Class. Quantum Grav., \textbf{29}, 245014 (2012).

\bibitem{Anagnost/2021} F. K. Anagnostopoulos, S. Basilakos, E. N. Saridakis, Phys. Lett. B, \textbf{822}, 136634 (2021).
 
\bibitem{Capo/2022} S. Capozziello, R. D'Agostino, Arxiv: 2204.01015 (2022).

\bibitem{Gruber/2014} C. Gruber, O. Luongo, Phys. Rev. D, \textbf{89}, 103506 (2014).

\bibitem{Capo/2020} S. Capozziello, R. D'Agostino, O. Luongo, Mon. Not. Roy. Astron. Soc., \textbf{494}, 2576 (2020).


\end{thebibliography}
\end{document}